\documentclass[a4paper,fleqn,usenatbib,useAMS]{mnras}

\usepackage{newtxtext,newtxmath}

\usepackage[T1]{fontenc}
\usepackage{ae,aecompl}

\usepackage{graphicx}	

\usepackage{amsmath}	

\usepackage{amssymb}	

\usepackage{wasysym}
\usepackage{color}
\usepackage{epsfig}
\usepackage{longtable}
\usepackage{pdflscape}
\usepackage{mathtools}
\usepackage{float}
\usepackage{footnote}
\usepackage{epstopdf}
\usepackage{units}
\usepackage{enumerate}
\usepackage{pifont}

\DeclareRobustCommand{\ion}[2]{\textup{#1\,\textsc{\lowercase{#2}}}}
\DeclareRobustCommand{\rgc}{R_{\rm GC}}
\DeclareRobustCommand{\teff}{T_{\mathrm{eff}}}
\DeclareRobustCommand{\muleo}{$\mu$-Leo}
\newcommand{\cmark}{\ding{51}}%

\newcommand{\xmark}{\ding{55}}%

\title[OCCASO III. Iron peak and $\alpha$ elements of 18 open clusters]{OCCASO III. Iron peak and $\alpha$ elements of 18 open clusters. Comparison with chemical evolution models and field stars.}

\author[L. Casamiquela et al.]{
L. Casamiquela$^{1}$\thanks{E-mail: laia.casamiquela-floriach@u-bordeaux.fr}, S. Blanco-Cuaresma$^{2}$, R. Carrera$^{3}$, L. Balaguer-N\'u\~nez$^{4}$,\newauthor
C. Jordi$^{4}$, F. Anders$^{4}$, C. Chiappini$^{5}$, J. Carbajo-Hijarrubia$^{4}$, D. S. Aguado$^{6}$,\newauthor A. del Pino$^{7}$, L. D\'iaz-P\'erez$^{8,9}$, C. Gallart$^{8}$, E. Pancino$^{10,11}$
\\
$^{1}$Laboratoire d'Astrophysique de Bordeaux, Univ. Bordeaux, CNRS, B18N, all\'ee Geoffroy Saint-Hilaire, 33615 Pessac, France\\
$^{2}$Harvard-Smithsonian Center for Astrophysics, 60 Garden Street, Cambridge, MA 02138, USA\\
$^{3}$INAF-Osservatorio Astronomico di Padova, vicolo dell’Osservatorio 5, 35122 Padova, Italy\\
$^{4}$Departament de F\'isica Qu\`antica i Astrof\'isica, Universitat de Barcelona, IEEC-UB, Mart\'i i Franqu\`es 1 08028 Barcelona, Spain\\
$^{5}$Leibniz-Institut fur Astrophysik Potsdam (AIP), An der Sternwarte 16, 14482 Potsdam, Germany\\
$^{6}$Institute of Astronomy, University of Cambridge, Madingley Road, Cambridge CB3 0HA, UK\\
$^{7}$Space Telescope Science Institute, 3700 San Martin Drive, Baltimore, MD 21218, USA\\
$^{8}$Instituto de Astrof\'isica de Canarias, La Laguna, 38205 Tenerife, Spain\\
$^{9}$Departamento de Astrof\'isica, Universidad de La Laguna, 38207 Tenerife, Spain\\
$^{10}$INAF - Osservatorio Astrofisico di Arcetri, 50125 Firenze, Italy \\
$^{11}$ASI Science Data Center, 00133 Roma, Italy
\\
}

\date{Accepted XXX. Received YYY; in original form ZZZ}

\pubyear{2018}

\begin{document}
\label{firstpage}
\pagerange{\pageref{firstpage}--\pageref{lastpage}}
\maketitle

\begin{abstract}
The study of open-cluster chemical abundances provides insights on stellar nucleosynthesis processes and on Galactic chemo-dynamical evolution. In this paper we present an extended abundance analysis of 10 species (Fe, Ni, Cr, V, Sc, Si, Ca, Ti, Mg, O) for red giant stars in 18 OCCASO clusters. This represents a homogeneous sample regarding the instrument features, method, line list and solar abundances from confirmed member stars. We perform an extensive comparison with previous results in the literature, and in particular with the \emph{Gaia} FGK Benchmark stars Arcturus and \muleo. We investigate the dependence of [X/Fe] with metallicity, Galactocentric radius ($6.5<\rgc<11$ kpc), age ($0.3<Age<10$ Gyr), and height above the plane ($|z|<1000$ pc). We discuss the observational results in the chemo-dynamical framework, and the radial migration impact when comparing with chemical evolution models. We also use APOGEE DR14 data to investigate the differences between the abundance trends in $\rgc$ and $|z|$ obtained for clusters and for field stars.\\
\end{abstract}

\begin{keywords}
techniques: spectroscopic -- Galaxy: open clusters and associations: general -- Galaxy: disc

\end{keywords}

\section{Introduction}
Open Clusters (OCs) are gravitationally-bound groups of stars formed from the same cloud of gas and dust, having the same age, distance, motion and initial chemical composition. Their ages cover the entire lifespan of the Galactic disc, tracing the young to old thin disc components. OCs are therefore widely used in the study of a variety of topics from stellar physics and understanding of the initial mass function to the process of assembly and evolution of the Galaxy. Also, OCs spanning different ages and chemical compositions are perfect targets to calibrate and validate astrometric, photometric and spectroscopic surveys.

The evolution of the chemical abundance gradients in the disk constitute an important constraint to the chemo-dynamical models \citep[e.g.][and references therein]{Anders+2017,Magrini+2017}.
The chemical gradients across the Galactic disk and the dependence of element abundances with age, are some of the most used observables to constrain the Galaxy disk evolution models. OCs are very valuable targets in these studies because they provide reliable ages and distances, even though the range in age is limited by the disruption of older clusters in the Galactic disk. Aside of providing insights in the chemical evolution of the thin disk, observed chemical abundances are a consequence of the nucleosynthetic processes, supernova yields, mechanisms to spread the produced elements, and star formation history of the Galaxy. The iron gradient have been analyzed in large samples of OCs by different authors \citep[e.g. ][]{Twarog+1997,Friel+2002,Frinchaboy+2013,Cantat-gaudin+2016,Jacobson+2016}, generally obtaining a decreasing trend of [Fe/H] towards large Galactocentric radii. However, few works study the spatial and temporal distribution of individual chemical species in OCs \citep[e.g. ][]{Yong+2012,Carrera+2011,Pancino+2010}, and only \citet{Magrini+2017} and \citet{Donor+2018} investigate this for a large homogeneous data set.

The Open Clusters Chemical Abundances from Spanish Observatories survey \citep[OCCASO, see][hereafter referred to as Paper I, for a detailed description]{Casamiquela+2016} is currently obtaining high-resolution spectra ($R\gtrsim$65,000) in the optical range (5000-8000\,\AA) for Red Clump (RC) stars in Northern hemisphere OCs. This survey systematically targets OCs with at least six RC stars per cluster, with a signal-to-noise (S/N) around 70. It was designed to obtain precise radial velocities and detailed chemical abundances in OCs to analyze their kinematics and chemical trends in the Galactic disk. 

Several large spectroscopic surveys like Gaia-ESO survey \citep[GES][]{Gilmore+2012}, GALAH \citep{Martell+2017}, and the forthcoming 4MOST \citep{deJong+2016} and MOONS \citep{Cirasuolo+2011} surveys operate in the South and include OCs in their samples. APOGEE \citep{Majewski+2017} is sampling both hemispheres in the near infrared H-band, while WEAVE \citep{Dalton+2012} will operate in the North from next year. All these surveys, but UVES observations by GES, have a medium spectral resolution, $R < 25,000$, and a limited wavelength coverage.
The high spectral resolution is necessary to get typical accuracies of $\lesssim0.05$ dex, needed for chemical tagging \citep{Holtzman+2018}. A large wavelength coverage allows to investigate more lines of a given element, and more chemical elements such as neutron-capture elements with few visible lines.

Up to now, for most of the OCs studied in the literature, less than 5 members have been observed (although for some cases it can amount to 200). It is known that stars of the same cluster at different evolutionary stages may show different abundances due to diffusion or mixing \citep[e.g.][]{Souto+2018}. Therefore, to study the chemical evolution of the Galaxy it is needed to sample stars at the same evolutionary stage. RC stars are suitable for this type of analysis because after the turn-off the material is brought towards the surface, which for most elements means a recovery of the initial abundances \citep{Dotter+2017}.

The OCCASO survey provides a sample of high resolution radial velocities and detailed abundances in Northern hemisphere OCs. Our data naturally complements what has been done by GES, more focused in the inner disk OCs, and by APOGEE, in the infrared and with a lower resolution. We have done an accurate selection of the RC member stars using Gaia astrometry and photometry. The high spectral resolution used, large wavelength range covered, and common analysis strategy helps to ensure high precision in the retrieved abundances.

This is the fourth paper devoted to analyze OCCASO data. In Paper I \citep{Casamiquela+2016} we obtained radial velocities and membership selection for 77 stars in 12 clusters. With more observations, 18 clusters were analyzed in \citet[][ hereafter referred to as Paper II, for a detailed description]{Casamiquela+2017} with atmospheric parameters and iron abundances derived for 115 stars. In \citet{Casamiquela+2018} we analyzed in detail the $\alpha$ elements and the kinematics of the cluster NGC~6705 (M~11).
Here we present the analysis of iron peak (Fe, Ni, Cr, Sc, V) and $\alpha$ elements (Si, Ca, Ti, Mg, O) for the full sample of observed stars up to August 2016: 139 spectra (115 stars) in 18 clusters and 2 \textit{Gaia} FGK Benchmark Stars \citep[GBS, ][]{Heiter+2015}. The results derived here allow the investigation of the chemical distribution of these elements as a function of age, Galactocentric radius and height above the Galactic plane. Most of our clusters are outside of the solar radius, so the gradients in the inner disk are not tested.

The paper is organized as follows: general characteristics of the data and the reduction is explained in Sect.~2, the spectroscopic analysis including the choice of atmospheric parameters, chemical abundance determination methods, solar abundance scale and line list is detailed in Sect.~3, the results of the chemical abundance analysis of the OCs are explained in Sect.~4, where we also include the comparison with the literature of the GBS, and on a cluster-by-cluster basis. A description of the chemical patterns with Galactocentric radius, height above the Galactic plane and age is described in Sect.~5. Finally, general conclusions are included in Sect.~6.

\section{Observations and data reduction}\label{sec:obs}
OCCASO observations are performed with the high-resolution echelle spectrographs available at Spanish observatories: CAFE at the 2.2 m telescope in the Centro Astron\'omico Hispano en Andaluc\'ia (CAHA), FIES at the 2.5 m NOT telescope in the Observatorio del Roque de los Muchachos (ORM) and HERMES at the 1.2 m Mercator telescope also in the ORM. These instruments have similar resolution $R\geq 65,000$ and wavelength range coverage $4000\leq \lambda \leq 9000$ \AA. The typical obtained S/N is around 70 per pixel.

In this work we include data from the observational runs described in Paper I and II (81 nights of observations between January 2013 - August 2016) that include data for 115 stars in 18 OCs, and Arcturus and \muleo, two reference stars \citep[GBS][]{Heiter+2015}, observed in OCCASO for the sake of comparison.

The general properties of the 18 OCs are summarized in Table~\ref{tab:clusters}, where positions, distances and ages have been updated according to the \textit{Gaia} DR2 results \citep{Cantat-gaudin+2018,Bossini+2019}.

\begin{table}
  \caption{\label{tab:clusters}Clusters of OCCASO completed by the end of August 2016. Distance from the Sun $D$ is taken from \citet{Cantat-gaudin+2018}, uncertainties are of the order 1-10 pc.  $\rgc$, $z$ are computed assuming the Sun Galactocentric radius 8.34 kpc.} Ages are from the indicated references. We also list the $V$ magnitude of the Red Clump, [Fe/H] computed in Paper II, and the number of stars observed.
\setlength\tabcolsep{3.5pt}
\begin{centering}
\begin{tabular}{cccccccc}
\hline
Cluster & $D$   & $\rgc$ & $z$  & Age   & $V_{\text{RC}}$  & [Fe/H]$_{\mathrm{EW}}$ &Num. \\
        & (kpc) & (kpc)  & (pc) & (Gyr) & & & Stars \\
\hline
IC 4756  & 0.47 & 7.97 & +43  & 1.0$^\text{e}$ & 9    &    0.00 & 8 \\
NGC 188  & 1.86 & 9.39 & +709 & 4.9$^\text{e}$ & 12.5 &    0.03 & 6 \\
NGC 752  & 0.44 & 8.64 & -174 & 1.5$^\text{e}$ & 9    &    0.01 & 7 \\
NGC 1817 & 1.72 & 10.01& -389 & 1.1$^\text{a}$ & 12.5 &   -0.09 & 5 \\
NGC 1907 & 1.56 & 9.89 & +9   & 0.4$^\text{b}$ & 9    &   -0.04 & 6 \\
NGC 2099 & 1.44 & 9.77 & +77  & 0.4$^\text{c}$ & 12   &    0.08 & 7 \\
NGC 2420 & 2.55 & 10.65& +858 & 1.9$^\text{e}$ & 12.5 &   -0.10 & 7 \\
NGC 2539 & 1.28 & 9.14 & +246 & 0.7$^\text{d}$ & 11   &    0.07 & 6 \\
NGC 2682 & 0.86 & 8.94 & +454 & 3.6$^\text{e}$ & 10.5 &    0.03 & 8 \\
NGC 6633 & 0.39 & 8.03 & +57  & 0.8$^\text{e}$ & 8.5  &    0.03 & 4 \\
NGC 6705 & 2.20 & 6.47 & -106 & 0.3$^\text{f}$ & 11.5 &    0.17 & 8 \\
NGC 6791 & 4.53 & 8.00 & +857 & 8.5$^\text{e}$ & 14.5 &    0.22 & 6 \\
NGC 6819 & 2.60 & 8.02 & +383 & 2.0$^\text{e}$ & 13   &    0.08 & 6 \\
NGC 6939 & 1.87 & 8.72 & +397 & 1.3$^\text{g}$ & 13   &    0.10 & 6 \\
NGC 6991 & 0.56 & 8.33 & +15  & 1.3$^\text{h}$ & 10   &    0.00 & 6 \\
NGC 7245 & 3.31 & 9.56 & -107 & 0.4$^\text{i}$ & 13   &    0.05 & 6 \\
NGC 7762 & 0.97 & 8.82 & +99  & 2.5$^\text{j}$ & 12.5 &    0.02 & 6 \\
NGC 7789 & 2.07 & 9.42 & -194 & 1.8$^\text{a}$ & 13   &    0.04 & 7 \\
\hline
\end{tabular}
\end{centering}

$^\text{a}$\citet{Salaris+2004}; $^\text{b}$\citet{Subramaniam+1999}; $^\text{c}$\citet{Nilakshi+2002}; $^\text{d}$\citet{Vogel+2003}; $^\text{e}$\citet{Bossini+2019}; $^\text{f}$\citet{cantatgaudin+2014b}; $^\text{g}$\citet{Andreuzzi+2004}; $^\text{h}$\citet{Karchenko+2005}; $^\text{i}$\citet{Subramaniam+2007}; $^\text{j}$\citet{Carraro+2016}; $^\text{k}$\citet{Krusberg+2006}.\\
\end{table}

The data reduction strategy is fully explained in Paper II. We built a dedicated pipeline which uses the results from the wavelength calibration of the instrument pipelines, and performs skyline subtraction, telluric correction, normalization and order merging.

\section{Spectroscopic analysis}\label{sec:analysis}
In this section we detail the analysis strategy concerning the line list and atmosphere models used, the atmospheric parameters adopted, the derivation of the chemical abundances, and the solar abundance scale. 

\subsection{Atmosphere models and line list}\label{sec:linelist}
We adopted the MARCS grid\footnote{\url{http://marcs.astro.uu.se/}} of spherically-symmetric model atmospheres of \citet{Gustafsson2008}, which assume the Solar abundances of \citet{Grevesse+2007} and $\alpha-$enhancement at low metallicities.

The master line list used in OCCASO is the Gaia-ESO Survey one \citep{Heiter+2015b}. It covers a wavelength range between $4200 \le \lambda \le 9200$ \AA. It contains atomic information for 35 different chemical species including the ones analyzed in this work.

For the chemical species \ion{Ni}{I}, \ion{Cr}{I}, \ion{Si}{I}, \ion{Ca}{I}, \ion{Ti}{I} we used the Equivalent Width (EW, see next subsection) analysis method, with a pre-selection of the Gaia-ESO master line list. We select lines that provide consistent abundances among our stars, and we reject lines with blends or with bad atomic parameters. This process is explained in detail in Paper II, but basically it rejects lines which give systematically discrepant abundances with respect to the average abundance of the chemical species. The cleaned line list is detailed in Table~\ref{tab:linelist}.

For \ion{V}{I}, \ion{Sc}{II}, \ion{Mg}{I} and \ion{O}{I}, abundances were obtained using spectral Synthesis (SS, see next subsection) method. The lines for these elements were selected to give consistent abundances, and are also included in Table~\ref{tab:linelist}.

To determine the Mg abundances we used three lines at 5528.405, 5711.088, and 6318.717 \AA. Their hyperfine structure splitting was taken into account in the line list. We used the mean of the three lines to derive the overall Mg abundance per spectrum. For oxygen we use the forbidden [\ion{O}{I}] line at 6300.304~\AA. This line is blended with a \ion{Ni}{I} line \citep{AllendePrieto+2001}, and this was taken into account to perform an accurate fit with the synthetic spectra. It is possible to measure oxygen abundances from the O triplet at $7774\,\AA$, but we do not attempt to use them because these lines are affected by very large NLTE effects.

\begin{table}
\begin{centering}
  \small
  \centering
  \caption{\label{tab:linelist}\ion{Ni}{I}, \ion{Cr}{I}, \ion{Si}{I}, \ion{Ca}{I}, \ion{Ti}{I}, \ion{V}{I}, \ion{Sc}{II}, \ion{Mg}{I}, \ion{O}{I} lines within master line list. Excitation potential $\chi$, and oscillator strengths $\log gf$ are indicated. References for the $\log gf$ are listed in the last column. The complete version of the table is available online. Here only the first lines are shown.} 
  \setlength\tabcolsep{3.1pt}
\begin{tabular}{ccccc}
\hline
$\lambda$ (\AA) & Element& $\chi\,\left( \mathrm{eV} \right)$ & $\log gf$ & Ref$^{\star}$ \\
\hline
5528.405 & \ion{Mg}{I} & 4.346 & -0.620 & KU \\
5711.088 & \ion{Mg}{I} & 4.350 & -1.620 & J05 \\
6318.717 & \ion{Mg}{I} & 5.108 & -2.020 & KU \\
6300.304 & [\ion{O}{I}]& 0.0   & -9.717 & C08    \\
5645.613 & \ion{Si}{I} & 4.93  &-2.043  & GARZ|BL \\
5665.555 & \ion{Si}{I} & 4.92  &-1.940  & GARZ|BL \\
5261.704 & \ion{Ca}{I} & 2.521 & -0.579 & SR  \\
5349.465 & \ion{Ca}{I} & 2.709 & -0.310 & SR  \\
5512.980 & \ion{Ca}{I} & 2.933 & -0.464 & S   \\
\hline
\end{tabular}
\flushleft $^{\star}$When two references separated by comma are listed, it means that the mean value of the $\log gf$ is taken. When two references separated by "|" are listed, it means that relative gf-values from the first source were re-normalised to an absolute scale using accurate lifetime measurements from the second source (Heiter et al., in preparation).\\

\flushleft References. B82: \citet{1982MNRAS.199...21B}, G89: \citet{1989A&A...208..157G}, B83: \citet{1983MNRAS.204..883B}, B86: \citet{1986MNRAS.220..289B}, N93: \citet{1993PhyS...48..297N}, L13: \citet{2013ApJS..205...11L}, W14: \citet{2014ApJS..211...20W}, GARZ: \citet{GARZ}, BL: \citet{BL}, GESMCHF: \citet{GESMCHF}, K07: \citet{K07}, K08: \citet{K08}, K10: \citet{K10}, LWST: \citet{LWST}, MFW: \citet{MFW}, NWL: \citet{NWL}, S: \citet{S}, SLS: \citet{SLS}, SR: \citet{SR}, KU: \url{http://cfaku5.cfa.harvard.edu/grids.html
}, J05: \citet{Johnson+2005}, C08: \citet{Caffau+2008}, J03: \citet{Johansson+2003}
\end{centering}
\end{table}

\subsection{Atmospheric parameters}\label{sec:AP}
Effective temperature ($\teff$), surface gravity ($\log g$) and iron abundances [Fe/H] were computed for all analysed stars in Paper II. In brief, atmospheric parameters were computed for all spectra using two different analysis methods, EW and SS (details on the two methodologies are given in Sec. 3.3). Both methods produce consistent results and we find internal differences compatible with the uncertainties. External tests were made using the Gaia FGK Benchmark Stars spectra, and comparing with determinations from photometry. Given the small differences found, we average the results of the $\teff$ and $\log g$ to perform the chemical analysis. Microturbulence ($\xi$) is let free when calculating the abundances with each method (SS and EW).
We refer to the Paper II for details in the procedure, and to their table~3 for the results used in this work.

\subsection{Chemical abundance determination}\label{sec:abund}
We use two methods (EW and SS) to compute chemical abundances, depending on the chemical species.

Chemical abundances of Ni, Cr, Si, Ca and Ti were obtained using the EW method. 
The EW analysis is performed in two steps. First, we use \texttt{DAOSPEC} \citep{Stetson+2008}, to measure EWs of the identified lines from a provided line list. Obtained EWs are the fed to \texttt{GALA} \citep{Mucciarelli+2013} that uses the plane-parallel radiative transfer code \texttt{WIDTH9} \citep{Kurucz2005} to derive chemical abundances.

V, Sc, Mg and O abundances are computed by SS fitting using \texttt{iSpec}
\citep{BlancoCuaresma+2014,BlancoCuaresma2019}, adapting a pipeline used in \citet{BlancoCuaresma+2018}. Spectral synthesis accounts for blends and hyperfine structure splitting present in the lines of those elements, and that cannot be taken into account using an EW method. The spectra are synthesized using the plane-parallel radiative transfer code SPECTRUM \citep{Gray+1994}, which is integrated in \texttt{iSpec}, using the atmospheric parameters determined in Paper II. The effective temperature, surface gravity and overall metallicity are kept fixed in a first run of \texttt{iSpec}, to constrain the optimal microturbulence velocity and spectral resolution to reproduce line shapes. After that, the region around the feature of interest is fitted to the synthetic spectra computed on the fly. 

\begin{figure}
\centering
\includegraphics[width=0.5\textwidth]{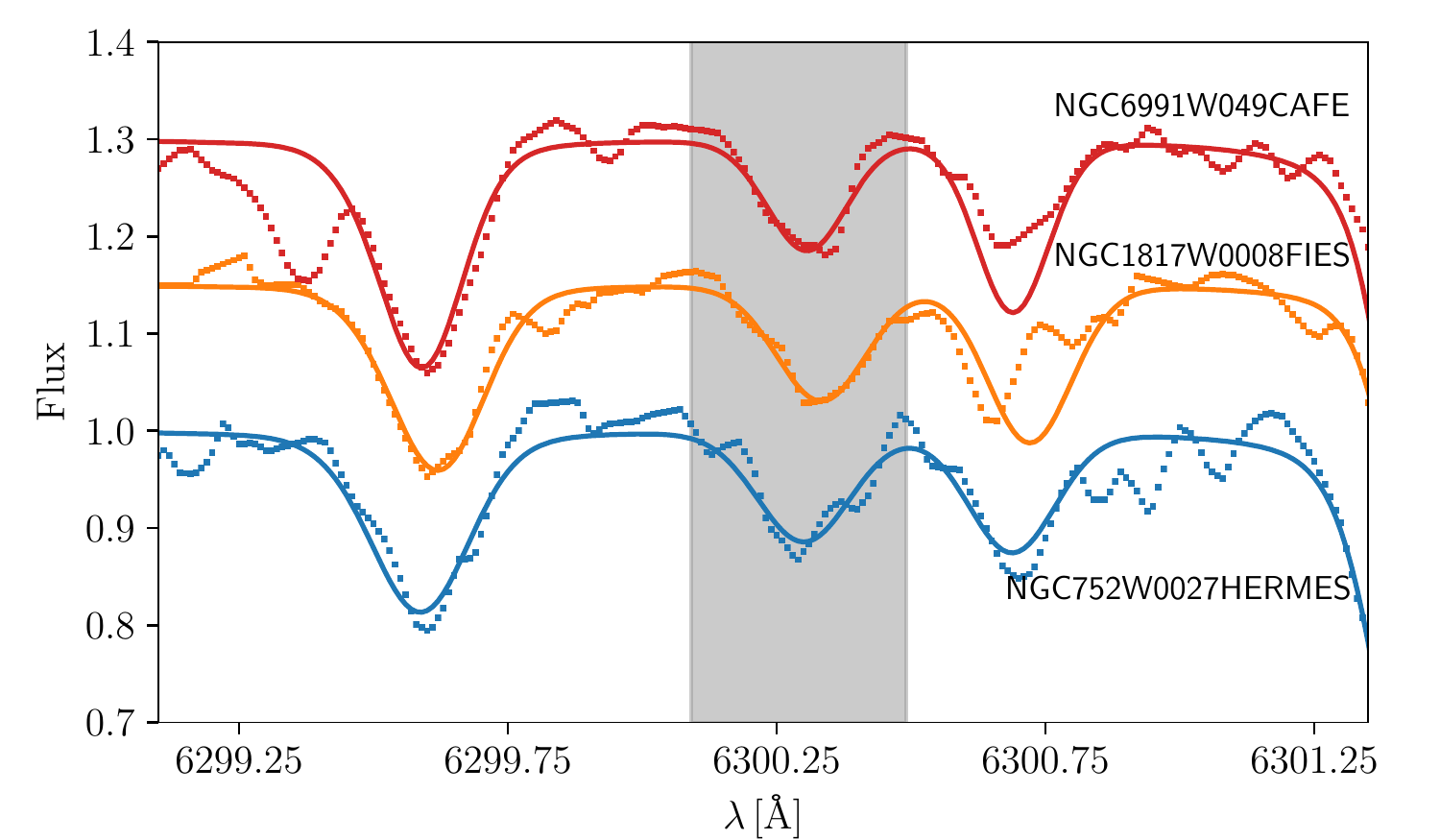}
\caption{Example of three synthetic fits of the [\ion{O}{I}] line at $6300.304$ \AA, used to retrieve oxygen abundances. The spectra correspond to three representative stars observed with the three instruments. Dotted lines are the observed spectra and solid lines are the best fit. The shaded region is where the fit is performed.}\label{fig:spectra_example}
\end{figure}

Three examples of the typical fits for the [\ion{O}{I}] are shown in Fig.~\ref{fig:spectra_example}. For this particular line, the fits to some of the spectra could not be performed or were discarded due to skyline contamination of low S/N in the surrounding spectral region.

We use the neutral transitions of all elements but Sc because we find larger number of lines. For Sc we use the lines from its single ionized stage \ion{Sc}{II} because there are more lines and their line by line abundances are more internally consistent (i.e. lower dispersion) than for neutral Sc. This has to be taken into account in comparisons with literature.

In Paper II we retrieved abundances of Fe using the same two methodologies. In this work we use the [Fe/H] derived from EW to calculate [X/Fe] for all elements analyzed.

\subsection{Solar abundance scale}

We have set our own solar abundance scale calculating abundances in the seven solar spectra available in the GBS high resolution spectral library \citep{Blanco+2014}. These spectra have been acquired with different instruments and all of them have been convolved to the OCCASO minimum resolution. We used the same line selection and model atmospheres as for the rest of OCCASO stars. We assumed the atmospheric parameters of the Sun derived in \citet{Heiter+2015}.

\begin{figure}
\centering
\includegraphics[width=0.5\textwidth]{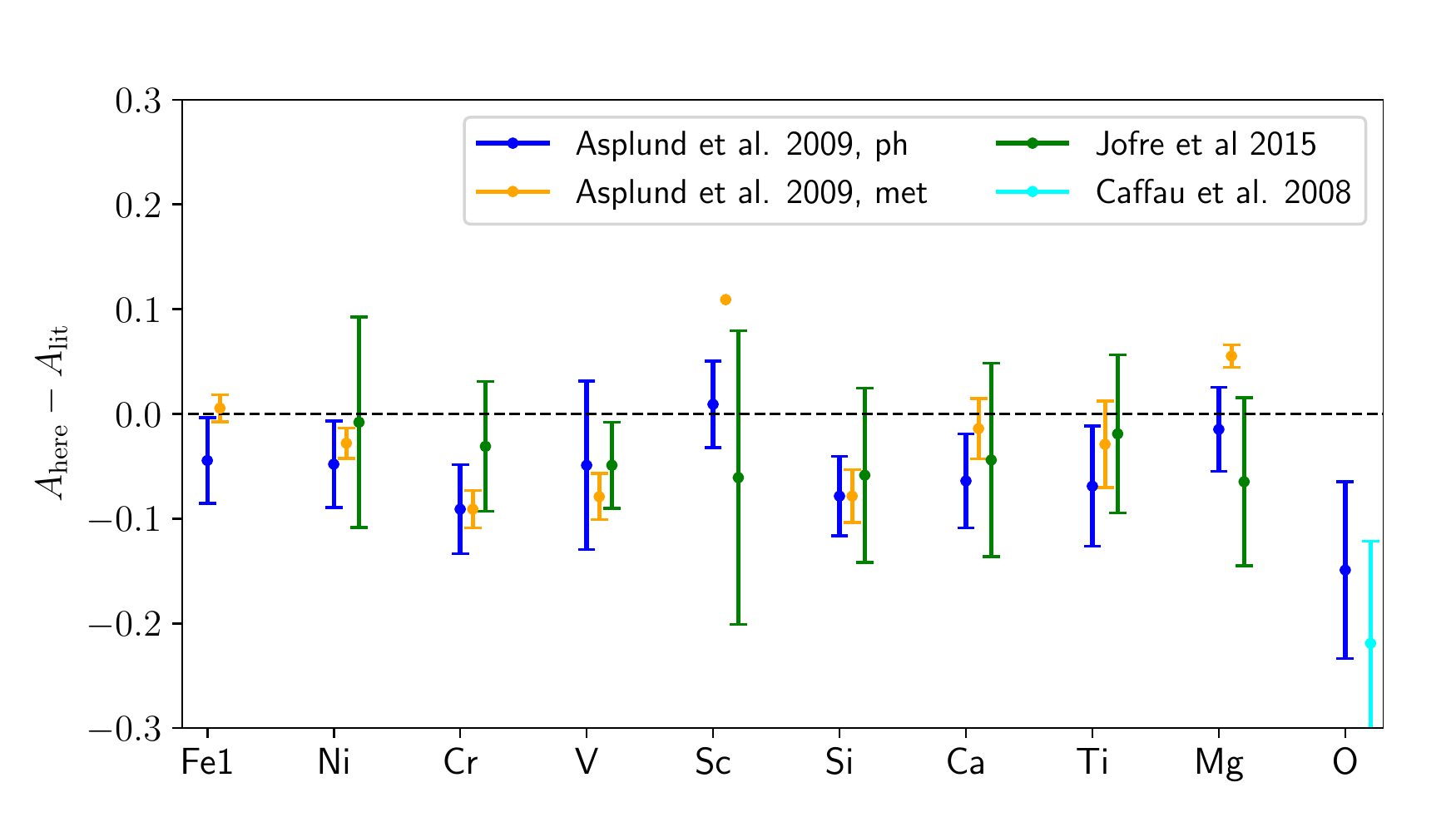}
\caption{Comparison of abundances [X/H] (here -- literature) obtained for the Sun with respect to previous works in the literature. For \citet{Asplund+2009} we compare with the photospheric and meteoritic abundances (except for [O/H]). Error bars represent the quadratic sum of the quoted errors by the two sources.}\label{fig:Sunlit}
\end{figure}

We have derived the absolute\footnote{Absolute abundance for a given element $A_X=\log (N_X/N_H)+12$, where $N_X$ and $N_H$ are the number of absorbers of the element X and of hydrogen, respectively.} solar abundances for each element as the median of the values of the $7$ spectra:
$A\left(\mathrm{Ni}\right)_{\odot}=6.17\pm0.01$,
$A\left(\mathrm{Cr}\right)_{\odot}=5.55\pm0.02$,
$A\left(\mathrm{V }\right)_{\odot}=3.88\pm0.01$,
$A\left(\mathrm{Sc}\right)_{\odot}=3.16\pm0.01$,
$A\left(\mathrm{Si}\right)_{\odot}=7.43\pm0.02$,
$A\left(\mathrm{Ca}\right)_{\odot}=6.28\pm0.02$,
$A\left(\mathrm{Ti}\right)_{\odot}=4.88\pm0.02$,
$A\left(\mathrm{Mg}\right)_{\odot}=7.58\pm0.01$,
$A\left(\mathrm{O }\right)_{\odot}=8.54\pm0.07$.
The quoted errors are computed as the median absolute deviation ($MAD$) of the $7$ values. These values are consistent within $1-2\sigma$ with previous determinations of the solar abundance scale, such as \citet{Asplund+2009} and \citet{Jofre+2015} (see Figure~\ref{fig:Sunlit}). O is the element which gives the largest differences with the literature, for both \citet{Asplund+2009} and \citet{Caffau+2008}. We cannot discard that there is some systematic difference in the absolute abundances of this element, but we have to notice that the comparison with literature of the bracket abundances for Arcturus and $\mu$Leo (see Sec. 4.1) and for clusters (see Sec. 4.3) show no systematics.

\section{Chemical abundance results}

\subsection{Gaia FGK Benchmark Stars}
\begin{figure}
\centering
\includegraphics[width=0.5\textwidth]{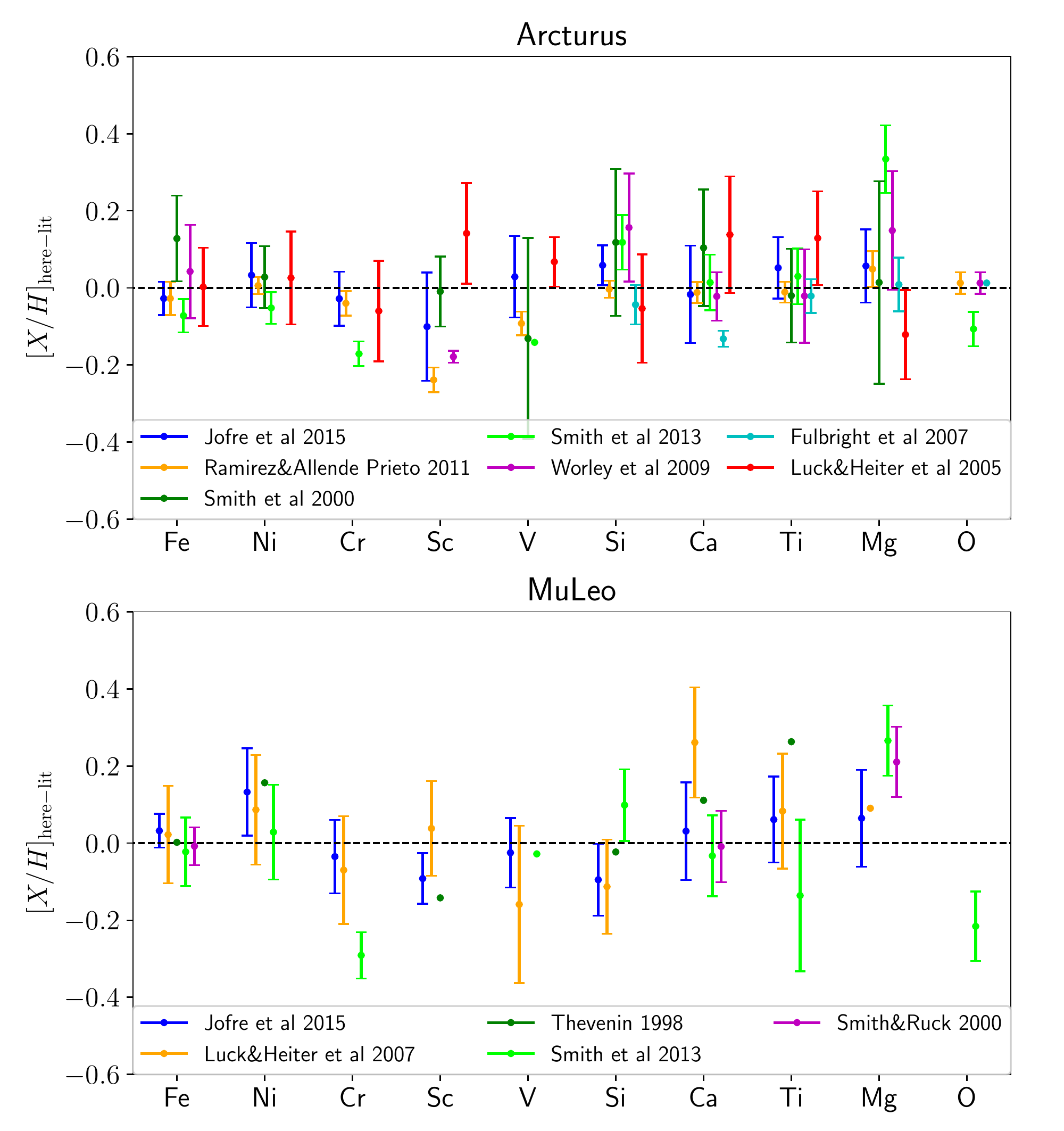}
\caption{Comparison of abundances [X/H] (here -- literature) obtained for Arcturus and \muleo$\,$ with respect to previous works in the literature. In the case of \citet{Smith+2013} we compare absolute abundances (not relative to the Sun).} Some values in the literature did not indicate any error bar.\label{fig:GBSlit}
\end{figure}

We have analysed in detail the elemental chemical abundances for the two GBS (Arcturus and \muleo) observed in OCCASO. Both stars were observed with the three instruments FIES, HERMES and CAFE. We computed the median abundances with respect to the Sun of the individual spectra equally weighted, and their $MAD$. We compare the obtained results with previous results in the literature \citep{Jofre+2015,Ramirez+2011,SmithV+2000,Smith+2013,Worley+2009,Fulbright+2007,Luck+2005,Luck+2007,Thevenin1998,Smith+2000} in Fig. ~\ref{fig:GBSlit}. Error bars in the figure are the quadratic sum of the errors of the OCCASO value ($MAD$ of the three instruments) and the literature error, if available.

The comparison shows a general good agreement within $1\sigma$ of the uncertainties. Average differences per author are always lower than 0.04 and 0.07, for Arcturus and \muleo, respectively. Difference spreads per author are lower than 0.14 for Arcturus and 0.16 for \muleo. We see the largest dicrepancies when we compare with the analysis in the H band done by \citet{Smith+2013}, particularly in the case of $\mu$Leo. This author provides only absolute abundances, and therefore we compare them with our absolute abundances. The cause of the discrepancies can be systematic differences in the absolute abundances, which are then mitigated by the computation of  bracket abundances\footnote{Notice for example that in Arcturus, computing bracket abundances [O/H] out of \citet{Smith+2013} value, using \citet{Ramirez+2011} solar abundance scale, leads also to a similar value of our [O/H] and of other literature values.}.

\subsection{Cluster abundances}\label{sec:clusterabund}
We have computed abundances with respect to the Sun of individual stars. The errors are calculated as the line-by-line spread in abundance divided by the square root of the number of lines. We have added quadratically the solar abundance error to compute errors in [X/H]. The values of [X/H] with their errors, per spectra are listed in Table~\ref{tab:finabundances}. We include in this table the results of all observed stars.

Some of the observed stars were already flagged as not bona fide members according to the analysis done in Paper II (NGC~188~W2051, NGC~1907~W2087, NGC~2539~W233, NGC~2682~W224, NGC~6791~W2604, NGC~6791~W3899, NGC~6939~W130, NGC~7245~W045, NGC~7762~W0084), and they were not used to compute cluster abundances. We have revisited the membership of all the stars according to Gaia DR2 memberships \citep{Cantat-gaudin+2018}. We have confirmed the previous discrepant cases, and we have attempted to classify between non-members (both astrometry and radial velocity/abundances not compatible) or spectroscopic binaries (compatible astrometry, not compatible radial velocity). Additionally, we have seen that IC~4756~W081 and NGC~7245~W205 do no appear in the membership analysis. IC~4756~W081 has a different (less than $3\sigma$) radial velocity with respect to the cluster mean, but compatible abundances. From the comparison with the literature of Paper I we suspect that it could be a binary and could have some problems in Gaia astrometry (large errors are quoted). NGC~7245~W205 has a slightly incompatible parallax but compatible proper motions, radial velocities and abundances. We consider both stars as members. A summary of the problematic stars is found in Table~\ref{tab:prob_stars}.

\begin{table}
\caption{Summary of the problematic stars detected as not members or spectroscopic binaries in Paper II (top part), and in Gaia DR2 (bottom part). We indicate if the star is compatible with the rest of the cluster from astrometry (proper motions $\mu$, parallax $\varpi$), radial velocities ($v_{\mathrm{r}}$) and abundances ($\mathrm{[X/H]}$). A $\sim$ means slightly compatible. A "-" in $\mathrm{[X/H]}$ means that abundances could not be computed. The membership that we derive is listed in the last column.}\label{tab:prob_stars}
\centering
\setlength\tabcolsep{2.1pt}
\begin{tabular}{|lccccccccccccc|}
 \hline
 OC & star & $\mu$, $\varpi$ & $v_{\mathrm{r}}$ & $\mathrm{[X/H]}$ & Memb$^{\star}$\\
 \hline
 NGC~188  & W2051 & \cmark & \xmark & \xmark & SB \\
 NGC~1907 & W2087 & \xmark & \xmark & \xmark & NM \\
 NGC~2539 & W233  & \cmark & \xmark & - & SB \\
 NGC~2682 & W224  & \xmark & \xmark & \cmark & NM \\
 NGC~6791 & W2604 & \cmark & \xmark & - & SB \\
 NGC~6791 & W3899 & \cmark & \xmark & \cmark & SB \\
 NGC~6939 & W130  & \cmark & \xmark & \xmark & SB \\
 NGC~7245 & W045  & \xmark & \cmark & \xmark & NM \\
 NGC~7762 & W0084 & \xmark & \xmark & \xmark & NM \\
 \hline
 IC~4756  & W081  & \xmark & $\sim$ & \cmark & M,SB? \\
 NGC~7245 & W205  & $\sim$ & \cmark & \cmark & M \\
 \hline\\
\end{tabular}

\noindent $^{\star}$ M$=$member, NM$=$non-member, SB$=$spectroscopic binary
\end{table}

Values of [X/Fe] per star are used to compute the cluster abundance ratios. In this case, the uncertainties are computed by quadratically adding errors in [X/H] and [Fe/H]. Cluster median [X/Fe] and $MAD$, computed only with the bona fide member stars, are listed in Table~\ref{tab:OCabundances}. For the stars observed with different instruments, a direct average was performed to obtain a final value per star and then compute the mean cluster abundance. In the case of O the number of stars is different because some spectra were rejected as explained in Sect. \ref{sec:abund}. As an indication of the precision in the element abundances we indicate in the last row of Table~\ref{tab:OCabundances} the median of the dispersions found for all clusters.

We plot the abundance ratio distributions of the OC stars as a function of their [Fe/H] abundances in appendix A. In general, clusters show homogeneous abundance patterns with few stars off the median value. We have quantified the total percentage of stars out of 2$MAD$, it ranges between 5 and 16\%, depending on the element. NGC~6791 has larger dispersions than the rest of the clusters for almost all analyzed elements, in particular for Ca, Si, Ti and Mg. This can be partly explained because the S/N of its spectra is lower than for the other OCs, since it is the faintest object in our sample. In the case of O, dispersions are larger than expected for some of the clusters (NGC~2099, NGC~2420) taking into account the spreads obtained for the other elements. We relate this to the intrinsic difficulty in analyzing the only measurable O line. In the case of NGC~7789 the O line has only been measured in a single star, so the value and uncertainty of this measurement is assigned to the cluster abundance.

\begin{figure}
\centering
\includegraphics[width=0.5\textwidth]{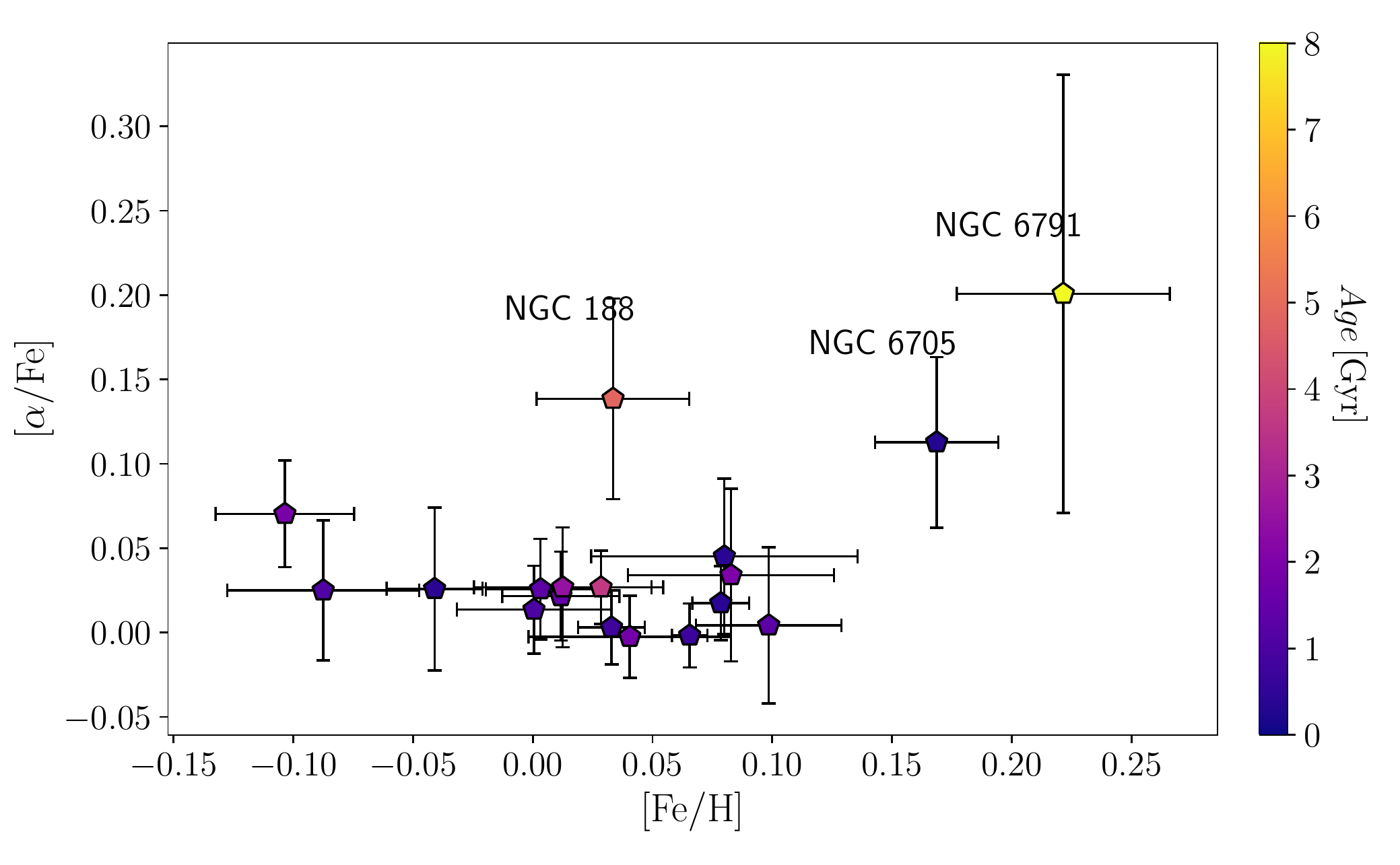}
\caption{Mean [$\alpha$/Fe] (Si, Ca, Ti, Mg) abundances as a function of [Fe/H] abundance, colored by age.}\label{fig:alphaFe}
\end{figure}

We plot [$\alpha$/Fe] computed as the mean of Si, Ca, Ti and Mg abundances as a function of [Fe/H] in Fig.~\ref{fig:alphaFe}. We have not used the O abundances in this computation because two of the clusters lack the oxygen value. In the plot we notice a slightly decreasing relation between $-0.1<$[Fe/H]$<0.1$. Three outliers stand out from the overall trend:

\begin{itemize}
 \item NGC~6791 is one of the oldest \citep[8.3 Gyr][]{Bossini+2019} and the most metal rich OC known, it has several [Fe/H] determinations: \citet{Donor+2018} $0.42\pm0.05$, \citet{Boesgaard+2015} $0.30\pm0.02$, \citet{Geisler+2012} $0.42\pm0.01$, \citet{Brogaard+2011} $0.29\pm0.03$, \citet{Gratton+2006} $0.47\pm0.04$, \citet{Carraro+2006} $0.39\pm0.01$, \citet{Friel+2002} $0.11\pm0.10$. In general, literature values are higher than ours ($0.22\pm0.04$\footnote{For this cluster very similar [Fe/H] is obtained using spectral synthesis (iSpec), $0.19\pm0.11$ see Paper II.}), except \citet{Friel+2002}. Consequently, literature abundance ratios differ with respect to our measures, particularly the $\alpha$ elements which we find supersolar compared to the roughly solar results by \citet{Donor+2018}, \citet{Boesgaard+2015}, \citet[][only for Ca, Ti and Si]{Carraro+2006}, and the very low [O/Fe]=-0.32 by \citet[][]{Gratton+2006}. Similar to our results, high $\alpha$ abundances were also found by \citet{Linden+2017}, based on APOGEE DR13 results. Given the age of the cluster and its chemical abundance,  \citet{Linden+2017} suggest this cluster has a thick disk origin, lying in the high metallicity tail of the high $\alpha$ sequence of the [$\alpha$/Fe] vs [Fe/H] plane.
 \item NGC~6705 is a young (300 Myr) OC for which three earlier works \citep{Magrini+2015,Magrini+2017,Casamiquela+2018} have found hints of $\alpha$ enhancement, particularly in Mg.
 \item NGC~188 is an old (6.3 Gyr) OC, with solar metallicity for which we find enhancement in [Si/Fe]=$0.09\pm0.03$ and [Mg/Fe]=$0.36\pm0.10$ using 5 stars, even though for two out of the five stars we could only measure two Mg lines (5711 and 6318 \AA). Slight overabundances in Si and Mg of the order of $\sim0.15$ were found in \citet{Friel+2010} and \citet{Jacobson+2011}.
\end{itemize}

\begin{table*}
 \caption{Fe, Ni, Cr, V, Sc, Si, Ca, Ti, Mg and O abundances for the stars (id from WEBDA$^{1}$ database) analyzed in OCCASO. The complete table is available as online data.}\label{tab:finabundances}
\footnotesize
\def\arraystretch{1.2}
\setlength\tabcolsep{2.1pt}
\centering
\begin{tabular}{|cccccccccccccc|}
 \hline 
  OC & star & Inst & [Fe/H]$^{\star}$ & [Ni/H] & [Cr/H] & [V/H] & [Sc/H] & [Ca/H] & [Si/H] & [Ti/H] & [Mg/H] & [O/H] \\
 \hline
 NGC1817 &  0008 & FIE & -0.11$\pm$0.01 &  -0.16$\pm$0.02 &  -0.13$\pm$0.02 &  -0.19$\pm$0.02 &  -0.14$\pm$0.04 &  -0.11$\pm$0.04 &  -0.07$\pm$0.03 &  -0.10$\pm$0.03 &  -0.08$\pm$0.11 &   0.06$\pm$0.07 \\
 NGC1817 &  0022 & FIE & -0.09$\pm$0.01 &  -0.14$\pm$0.03 &  -0.02$\pm$0.05 &  -0.16$\pm$0.02 &  -0.10$\pm$0.04 &  -0.07$\pm$0.09 &  -0.03$\pm$0.05 &  -0.07$\pm$0.03 &  -0.05$\pm$0.03 &  -0.13$\pm$0.07 \\
 NGC1817 &  0073 & FIE & -0.04$\pm$0.01 &  -0.06$\pm$0.02 &  -0.06$\pm$0.03 &  -0.09$\pm$0.01 &   0.00$\pm$0.04 &  -0.07$\pm$0.02 &   0.03$\pm$0.03 &  -0.00$\pm$0.04 &  -0.06$\pm$0.08 &  -0.12$\pm$0.07 \\
 NGC1817 &  0079 & FIE & -0.05$\pm$0.01 &  -0.06$\pm$0.03 &  -0.06$\pm$0.03 &  -0.11$\pm$0.02 &  -0.07$\pm$0.04 &   0.03$\pm$0.05 &   0.00$\pm$0.04 &  -0.01$\pm$0.04 &  -0.09$\pm$0.03 &   - \\
 NGC1817 &  0127 & FIE & -0.09$\pm$0.01 &  -0.08$\pm$0.03 &  -0.06$\pm$0.04 &  -0.10$\pm$0.02 &  -0.03$\pm$0.03 &   0.07$\pm$0.04 &  -0.00$\pm$0.03 &  -0.04$\pm$0.04 &  -0.11$\pm$0.00 &  -0.17$\pm$0.07 \\
 NGC2099 &  007  & HER & 0.05$\pm$0.01  &   0.06$\pm$0.02 &   0.12$\pm$0.04 &  -0.00$\pm$0.02 &   0.04$\pm$0.04 &   0.09$\pm$0.04 &   0.08$\pm$0.03 &   0.08$\pm$0.03 &   0.05$\pm$0.04 &   0.02$\pm$0.07 \\
 NGC2099 &  016  & HER & 0.09$\pm$0.01  &   0.06$\pm$0.02 &   0.10$\pm$0.04 &   0.01$\pm$0.02 &   0.07$\pm$0.04 &   0.10$\pm$0.07 &   0.07$\pm$0.03 &   0.10$\pm$0.04 &   0.04$\pm$0.02 &   0.15$\pm$0.07 \\
 NGC2099 &  031  & HER & 0.15$\pm$0.01  &   0.10$\pm$0.03 &   0.17$\pm$0.03 &   0.08$\pm$0.02 &   0.11$\pm$0.05 &   0.19$\pm$0.05 &   0.15$\pm$0.03 &   0.16$\pm$0.03 &   0.10$\pm$0.04 &   - \\
 NGC2099 &  148  & HER & 0.08$\pm$0.01  &   0.06$\pm$0.02 &   0.03$\pm$0.03 &   0.01$\pm$0.01 &   0.06$\pm$0.04 &   0.06$\pm$0.06 &   0.13$\pm$0.04 &   0.11$\pm$0.03 &   0.12$\pm$0.05 &  -0.09$\pm$0.07 \\
 NGC2099 &  172  & HER & 0.06$\pm$0.01  &   0.10$\pm$0.02 &   0.09$\pm$0.04 &  -0.01$\pm$0.02 &   0.02$\pm$0.06 &   0.16$\pm$0.05 &   0.08$\pm$0.03 &   0.10$\pm$0.03 &   0.05$\pm$0.04 &  -0.03$\pm$0.07 \\
 NGC2099 &  401  & HER & 0.09$\pm$0.01  &   0.06$\pm$0.02 &   0.08$\pm$0.03 &   0.00$\pm$0.02 &   0.15$\pm$0.04 &   0.12$\pm$0.06 &   0.14$\pm$0.03 &   0.07$\pm$0.03 &   0.03$\pm$0.04 &   0.20$\pm$0.07 \\
 NGC2099 &  488  & HER & 0.07$\pm$0.01  &   0.04$\pm$0.02 &   0.11$\pm$0.03 &   0.00$\pm$0.02 &   0.10$\pm$0.03 &   0.11$\pm$0.04 &   0.11$\pm$0.03 &   0.09$\pm$0.03 &   0.08$\pm$0.02 &   0.01$\pm$0.07 \\
 \hline
\end{tabular}
\flushleft $^{\star}$[Fe/H] values come from Paper II.\\
$^{1}$\url{https://webda.physics.muni.cz/navigation.html}
\end{table*}
 
\begin{table*}
 \caption{Mean cluster abundance ratios using bona fide member stars. Errors correspond to the standard deviation of the star abundances. The number of stars used to calculate the mean cluster abundances is indicated in the second column. The number of stars used to calculate abundance of O is indicated in the last column (see text). To given an idea of the uncertainties per element, we list in the last row the median of the cluster dispersions (i.e. errors associated to cluster abundances) and its $MAD$.\label{tab:OCabundances}}
\footnotesize
\setlength\tabcolsep{1pt}
\centering
\begin{tabular}{|p{1.45cm}cccccccccccc|}
 \hline 
   Name &  N & [Fe/H] & [Ni/Fe] & [Cr/Fe] & [V/Fe] & [Sc/Fe] & [Ca/Fe]& [Si/Fe] & [Ti/Fe] & [Mg/Fe] & [O/Fe] & N$_{\mathrm{O}}$\\
 \hline
  IC4756 &    8 &   $0.00\pm0.03$ &  $-0.03\pm0.01$ &   $0.00\pm0.03$ &  $-0.03\pm0.03$ &  $-0.03\pm0.02$ &   $0.05\pm0.02$ &  $-0.00\pm0.01$ &  $0.04\pm0.03$ &  $-0.04\pm0.04$ &   $0.03\pm0.03$ &   7 \\
 NGC1817 &    5 &  $-0.09\pm0.04$ &  $-0.02\pm0.04$ &  $-0.01\pm0.01$ &  $-0.06\pm0.02$ &  $-0.01\pm0.03$ &   $0.02\pm0.07$ &   $0.05\pm0.02$ &  $0.04\pm0.03$ &  $-0.01\pm0.04$ &  $-0.06\pm0.03$ &   4 \\
  NGC188 &    5 &   $0.03\pm0.03$ &   $0.09\pm0.04$ &   $0.04\pm0.02$ &   $0.01\pm0.05$ &  $-0.01\pm0.04$ &   $0.04\pm0.06$ &   $0.09\pm0.03$ &  $0.06\pm0.05$ &   $0.36\pm0.10$ &   $0.05\pm0.03$ &   2 \\
 NGC1907 &    5 &  $-0.04\pm0.02$ &  $-0.037\pm0.003$ &   $0.01\pm0.08$ &  $-0.11\pm0.03$ &  $-0.08\pm0.09$ &   $0.04\pm0.03$ &   $0.03\pm0.04$ &  $0.03\pm0.02$ &   $0.01\pm0.10$ &     -           &   0 \\
 NGC2099 &    7 &   $0.08\pm0.01$ &  $-0.02\pm0.01$ &   $0.02\pm0.02$ &  $-0.07\pm0.01$ &  $-0.01\pm0.03$ &   $0.04\pm0.01$ &   $0.03\pm0.03$ &  $0.02\pm0.01$ &  $-0.02\pm0.04$ &  $-0.04\pm0.12$ &   6 \\
 NGC2420 &    7 &  $-0.10\pm0.03$ &   $0.04\pm0.05$ &   $0.01\pm0.07$ &   $0.00\pm0.02$ &   $0.10\pm0.04$ &   $0.05\pm0.02$ &   $0.07\pm0.02$ &  $0.10\pm0.03$ &   $0.07\pm0.05$ &   $0.30\pm0.14$ &   6 \\
 NGC2539 &    5 &   $0.07\pm0.01$ &  $-0.00\pm0.02$ &   $0.01\pm0.02$ &  $-0.05\pm0.01$ &   $0.01\pm0.02$ &   $0.03\pm0.01$ &  $-0.00\pm0.02$ &  $0.03\pm0.02$ &  $-0.07\pm0.02$ &  $-0.04\pm0.03$ &   4 \\
 NGC2682 &    7 &   $0.03\pm0.03$ &   $0.06\pm0.02$ &   $0.01\pm0.01$ &  $-0.03\pm0.03$ &  $-0.04\pm0.03$ &   $0.02\pm0.03$ &   $0.05\pm0.01$ &  $0.04\pm0.01$ &   $0.01\pm0.03$ &   $0.04\pm0.09$ &   4 \\
 NGC6633 &    4 &   $0.03\pm0.01$ &  $-0.026\pm0.008$ &  $-0.01\pm0.01$ &  $-0.05\pm0.01$ &  $-0.00\pm0.02$ &   $0.02\pm0.02$ &   $0.00\pm0.03$ &  $0.01\pm0.01$ &  $-0.02\pm0.03$ &   $0.13\pm0.08$ &   4 \\
 NGC6705 &    8 &   $0.17\pm0.03$ &   $0.08\pm0.02$ &   $0.02\pm0.07$ &  $-0.02\pm0.05$ &   $0.02\pm0.03$ &   $0.03\pm0.07$ &   $0.17\pm0.02$ &  $0.04\pm0.03$ &   $0.22\pm0.08$ &   $0.11\pm0.06$ &   6 \\
 NGC6791 &    5 &   $0.22\pm0.04$ &   $0.14\pm0.02$ &   $0.13\pm0.05$ &   $0.26\pm0.04$ &   $0.02\pm0.04$ &   $0.17\pm0.17$ &   $0.17\pm0.08$ &  $0.07\pm0.13$ &   $0.40\pm0.14$ &   $0.19\pm0.10$ &   3 \\
 NGC6819 &    6 &   $0.08\pm0.04$ &   $0.05\pm0.01$ &  $-0.01\pm0.03$ &  $-0.06\pm0.04$ &   $0.02\pm0.01$ &   $0.00\pm0.05$ &   $0.03\pm0.05$ &  $0.05\pm0.04$ &   $0.05\pm0.08$ &     -           &   0 \\
 NGC6939 &    5 &   $0.10\pm0.03$ &   $0.00\pm0.02$ &   $0.00\pm0.03$ &  $-0.11\pm0.01$ &  $-0.05\pm0.02$ &   $0.06\pm0.07$ &  $-0.02\pm0.05$ &  $0.04\pm0.04$ &  $-0.06\pm0.03$ &  $-0.16\pm0.11$ &   4 \\
 NGC6991 &    6 &   $0.00\pm0.02$ &  $-0.03\pm0.01$ &  $-0.02\pm0.03$ &  $-0.02\pm0.01$ &  $-0.01\pm0.02$ &   $0.05\pm0.06$ &   $0.02\pm0.02$ &  $0.05\pm0.01$ &  $-0.01\pm0.04$ &   $0.12\pm0.09$ &   4 \\
 NGC7245 &    5 &   $0.08\pm0.06$ &   $0.00\pm0.04$ &   $0.07\pm0.05$ &  $-0.01\pm0.04$ &   $0.01\pm0.03$ &   $0.08\pm0.08$ &   $0.01\pm0.03$ &  $0.08\pm0.07$ &   $0.00\pm0.01$ &   $0.10\pm0.07$ &   2 \\
  NGC752 &    7 &   $0.01\pm0.02$ &   $0.00\pm0.03$ &  $-0.01\pm0.02$ &  $-0.05\pm0.02$ &  $-0.00\pm0.02$ &   $0.05\pm0.05$ &   $0.02\pm0.01$ &  $0.04\pm0.01$ &  $-0.02\pm0.03$ &   $0.13\pm0.02$ &   7 \\
 NGC7762 &    5 &   $0.01\pm0.04$ &   $0.01\pm0.07$ &   $0.019\pm0.001$ &  $-0.06\pm0.04$ &  $-0.01\pm0.01$ &  $-0.01\pm0.03$ &   $0.02\pm0.01$ &  $0.03\pm0.03$ &   $0.07\pm0.07$ &   $0.18\pm0.07$ &   5 \\
 NGC7789 &    7 &   $0.04\pm0.04$ &  $-0.01\pm0.01$ &   $0.03\pm0.04$ &  $-0.06\pm0.03$ &   $0.02\pm0.01$ &   $0.03\pm0.03$ &  $-0.00\pm0.01$ &  $0.01\pm0.03$ &  $-0.05\pm0.03$ &   $0.07\pm0.07$ &   1 \\
 \hline
 Median dispersions& & $0.03\pm0.01$ & $0.02\pm0.01$ & $0.03\pm0.02$ & $0.03\pm0.01$ & $0.02\pm0.01$ & $0.04\pm0.02$ & $0.02\pm0.01$ & $0.03\pm0.01$ & $0.04\pm0.02$ & $0.07\pm0.06$ \\
 \hline
\end{tabular}
\end{table*}

\subsection{Comparison with literature}

\begin{figure}
\centering
\includegraphics[width=0.5\textwidth]{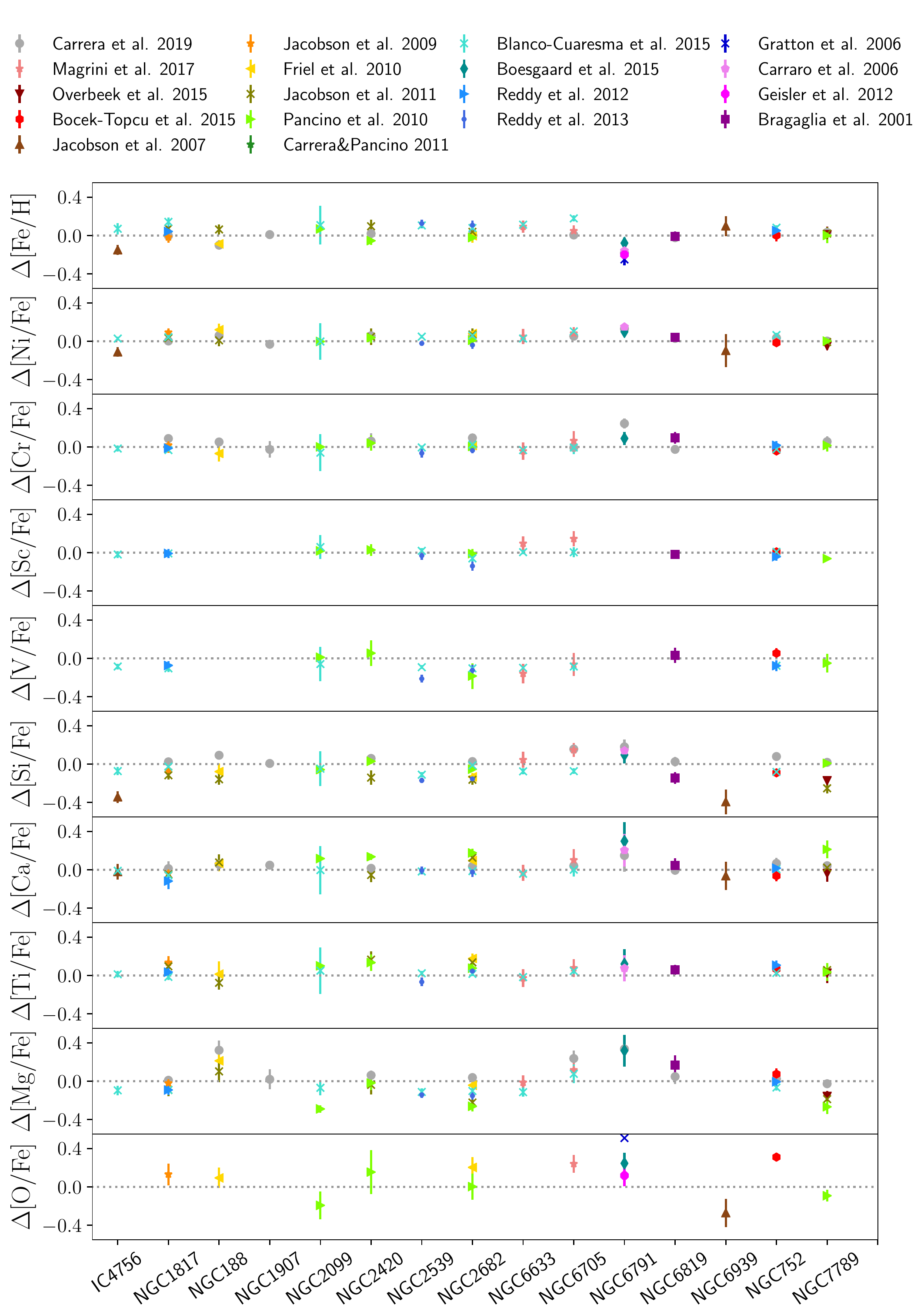}
\caption{Comparison of the obtained OC abundances with previous results in the literature (this work -- literature).}\label{fig:comp_lit}
\end{figure}

We compare the cluster abundances derived here with previous high-resolution abundances from the literature in Fig.~\ref{fig:comp_lit}.

For all elements we find mean offsets well within the dispersions: $ 0.01\pm0.09$ in [Fe/H], $ 0.03\pm0.05$ in [Ni/Fe], $ 0.01\pm0.06$ in [Cr/Fe], $ -0.07\pm0.07$ in [V/Fe], $ -0.00\pm0.06$ in [Sc/Fe], $-0.05\pm0.12$ in [Si/Fe], $ 0.03\pm0.08$ in [Ca/Fe], $ 0.05\pm0.06$ in [Ti/Fe], $ -0.01\pm0.15$ in [Mg/Fe], $ 0.11\pm0.20$ in [O/Fe]. The largest spread is found for the comparison of [O/Fe] for the intrinsic difficulty in computing its abundance.

We analyze in detail the comparison with \citet{Carrera+2019} because it is one of the most extensive and recent analysis of OCs abundances. This is a compilation of OC abundances obtained using the memberships from Gaia DR2 \citep{Cantat-gaudin+2018} and searching for stars in common with the APOGEE and GALAH latest results. The mean differences (us $-$ literature): $-0.02\pm0.06$ [Fe/H], $0.04\pm0.04$ [Ni/Fe], $0.05\pm0.07$ [Cr/Fe], $0.05\pm0.04$ [Si/Fe], $0.10\pm0.13$ [Mg/Fe]; are compatible and of the same order of the uncertainties in most of the provided elements. For the case of Mg we find a larger difference comparing with the other elements. The difference is anyway still inside the dispersion, showing that there is no systematics which affects the two data sets, only higher measurement uncertainties in probably both samples.

We obtain good agreement with previous values of clusters with well studied chemical signature, such as NGC~2682 (M~67) and NGC~7789.
The case of NGC~6791 is the most discrepant and it is already studied in detail in the previous subsection.

\subsection{Abundances of clusters and field stars from HARPS-GTO}

\begin{figure*}
\centering
\includegraphics[width=\textwidth]{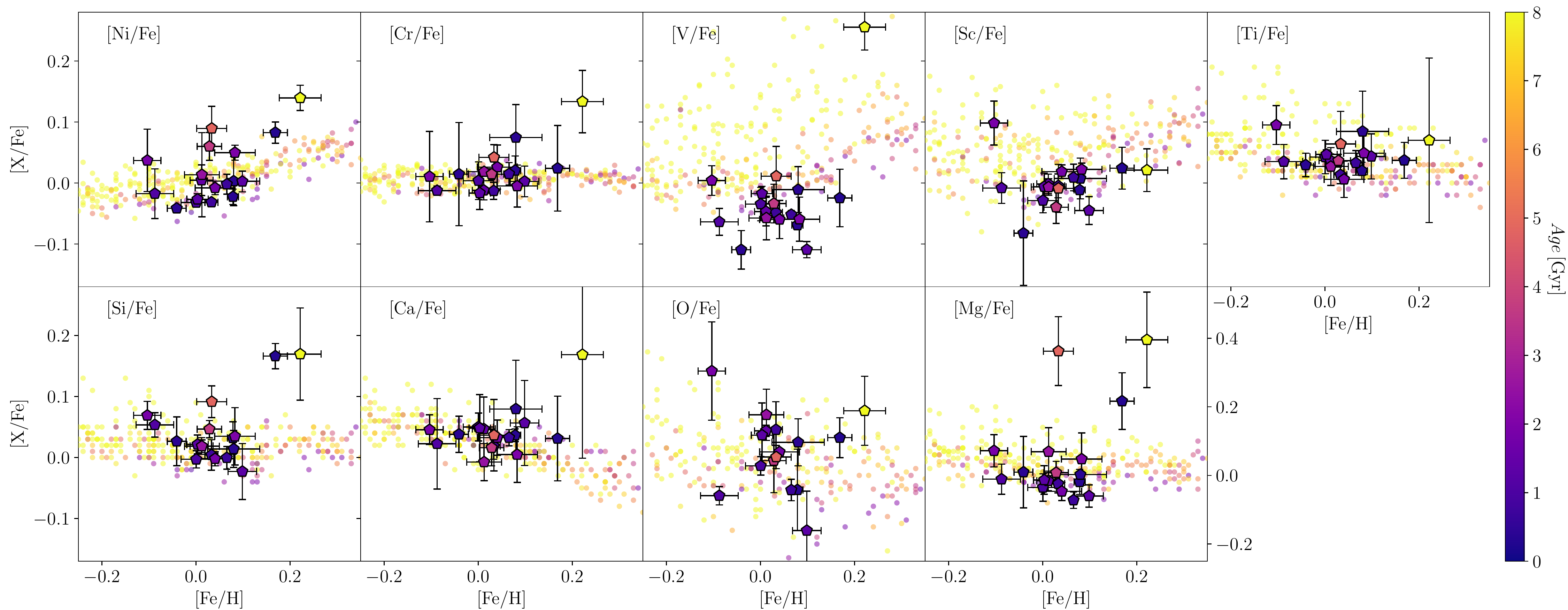}
\caption{OC mean [X/Fe] abundance ratios as a function of [Fe/H], color coded by age (pentagons with error bars). The small dots show the abundance trends found in the local disc, as traced by the HARPS-GTO stars with the same color code. Note the change of y scale in [Mg/Fe] and [O/Fe].}\label{fig:OC_XFe_FeH}
\end{figure*}

We have compared the abundance results and age distribution of the clusters with respect to the field stars using the HARPS-GTO sample. This is a sample of high quality spectra from dwarf field stars in the solar neighbourhood, used as a reference in a handful of works \citep[e.g. ][]{Adibekyan+2012,Anders+2018,DelgadoMena+2017,Minchev+2018}. We have used the recent reanalysis of the abundances and atmospheric parameters by \citet{DelgadoMena+2017}. Abundances of iron peak elements are taken from \citet{Adibekyan+2012}.
We have used the spectro-photometric ages determined by \citet{Anders+2018}, and we have applied the same quality cuts used in \citet{Minchev+2018}: $\delta\mathrm{[Mg/Fe]}<0.07$, $\delta Age<1$ Gyr or $\delta Age/Age<0.25$, and $5300<T_{\mathrm{eff}}<6000$ K. With this selection we obtain 397 stars. The cuts in effective temperature and age uncertainties restrict the age range to stars older than $\sim1.5$ Gyr. We plot both the HARPS-GTO stars and the OC abundances ratios with respect to Fe as a function of their [Fe/H], color coded by age in the Fig.~\ref{fig:OC_XFe_FeH}. The abundances of our clusters are the expected ones for the stars of the thin disk, so the distribution overlaps with that of the young field stars (blue-magenta points) in most of the element ratios. For V and O a more disperse picture is seen also for the field stars. NGC~6791 (the oldest cluster) is a clear outlier in the $\alpha$ elements Si, Mg and Ca, and in the Fe-peak elements Ni, Cr and V.

\section{Distribution of abundances in the Galactic disc}
Trends in the chemical abundances with Galactocentric radius and age provide valuable constraints on models of Galactic chemical evolution. A further constraint is the variation of these trends with time. The analysed OCs in OCCASO span a range in Galactocentric radius of $6.6<\rgc<11$ kpc, and have ages between $0.3$ and $10$ Gyr. However, there must be kept in mind that most of our clusters are concentrated between 8-10 kpc in Galactocentric radius, and mainly between 0.5-2 Gyr in age.
All the clusters in the sample have $|z|<1$ kpc (see Table~\ref{tab:clusters}). 

In the next subsections we analyze the distribution of abundances of OCCASO clusters with Galactocentric radius, distance above the Galactic plane and age. We compare with chemical evolution models and we complement our sample with field stars from APOGEE, and the most recent compilation of OCs abundances \citep{Carrera+2019} from APOGEE and GALAH.

\subsection{Abundances with $\rgc$}\label{sec:trends_rgc}

\begin{figure*}
\centering
\includegraphics[width=\textwidth]{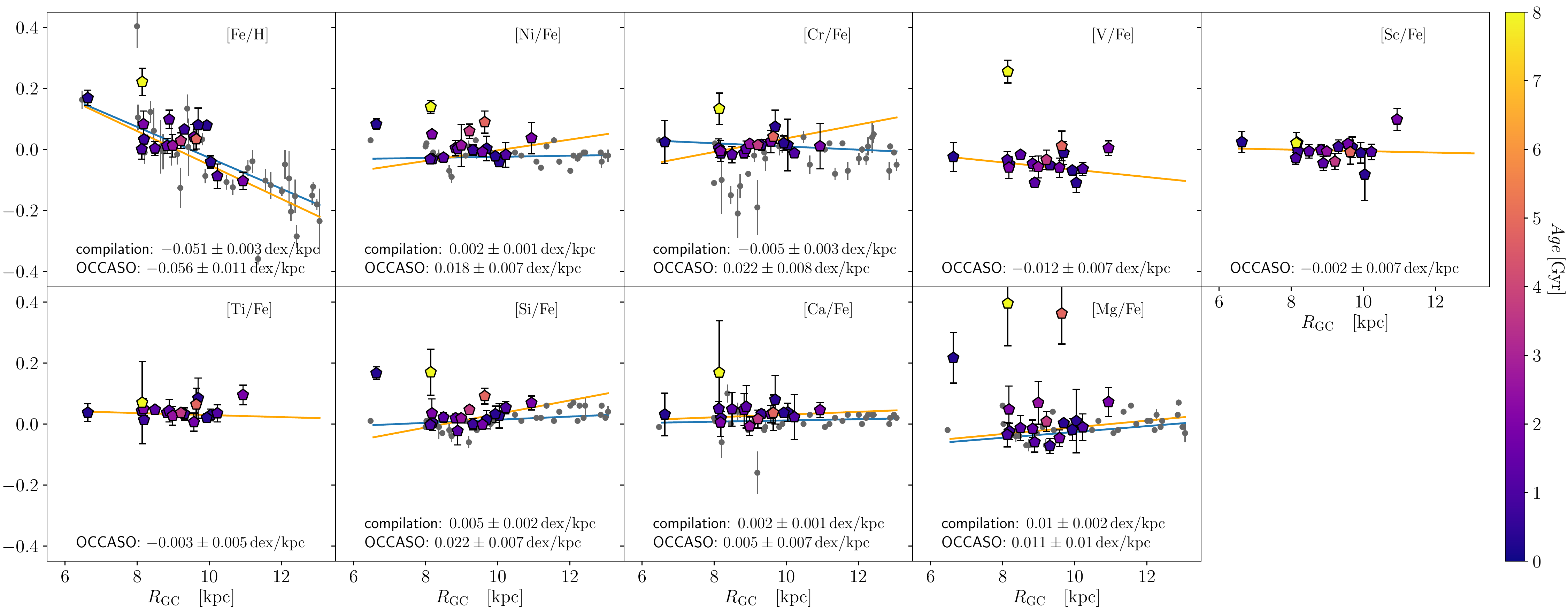}
\caption{OCCASO abundances [Fe/H] and [X/Fe] as a function of Galactocentric radius are represented as pentagons colored by age. Black circles are the OCs analyzed by \citet{Carrera+2019}. Linear fits using only OCCASO data (orange line) and the full compilation (blue line) are represented with the obtained slopes indicated in each panel.}\label{fig:abund_RGC}
\end{figure*}

We plot the abundances [Fe/H] and [X/Fe] of the different elements computed in Sec. 4 as a function of the Galactocentric radius in Fig.~\ref{fig:abund_RGC}. We overplot in this figure the OCs analyzed in \citet{Carrera+2019}. This is a compilation of OCs observed by APOGEE and GALAH, with abundances determined from four or more stars. Abundances for V, Sc, Ti and O (and cluster ages) are not available in this compilation because they are not completely reliable from APOGEE data \citep[as explained using literature comparisons by ][]{Jonsson+2018,Holtzman+2018}. A very good agreement between OCCASO and \citet{Carrera+2019} is already mentioned in the comparison done in Sec. 4.3. 
In this section we analyze the Galactocentric trends with the OCCASO sample only (18 clusters), and also with the two joint data sets (43 clusters) to improve statistics. For the seven clusters in common we take the mean of the two abundances and the mean of the errors as uncertainty.

We perform linear fits to the two samples to obtain the Galactic trends using a Bayesian outlier detector with a Markov Chain Monte Carlo (MCMC). We use the approach explained in detail in \citet{Hogg+2010}\footnote{General code available in AstroML webpage \url{http://www.astroml.org/book_figures/chapter8/fig_outlier_rejection.html}}. This method performs a linear regression coupled with an objective datapoint rejection, through a modelization of the outlier distribution. This approach is very convenient in our case, since the standard linear fitting is sensitive to outliers, particularly when the data is scarce. The method infers at the same time the two parameters of the linear fit, together with three parameters that define a distribution of outliers: mean, variance and fraction of bad points. The model is run through 25,000 MCMC realizations, and the best value of the slope is taken as the maximum of the posterior distribution.

To retrieve the present day metallicity gradient we have used only the youngest OCs from OCCASO. We obtain a decreasing linear trend in [Fe/H] as a function of $\rgc$ with slopes between $-0.05\pm0.01\,\mathrm{dex/kpc}$ and $-0.06\pm0.01\,\mathrm{dex/kpc}$, for age$<2$ Gyr (14 OCs) and age$<1.5$ Gyr (11 OCs), respectively. These are similar values to $-0.051\pm0.003\,\mathrm{dex/kpc}$, found by \citet{Genovali+2014} using Cepheids, showing that the clusters up to 1-2 Gyr trace the present day metallicity gradient because they have not migrated far from their birth place. For an extended discussion about this we refer to \citet{Anders+2017}.

Now we compute linear fits to all the chemical species using the OCCASO sample only. The slopes are indicated in the panels of Fig.~\ref{fig:abund_RGC}. In spite of the low number of clusters, OCCASO is a very homogeneous sample, in terms of analysis but also in terms of selection of stars in the cluster. In this case we are including OCs wih different ages, and the resulting gradients could be contaminated by the oldest clusters which most probably have moved form their birthplace, not only because of their orbits, but also because of the effect of radial migration. The slope in [Fe/H] is $-0.056\pm0.011\,\mathrm{dex/kpc}$, similar to the present day gradient, showing that the presence of old clusters do not affect much the sample, which is dominated by the youngest clusters.

Finally, we use the joint compilation of OCs (i.e. colored and grey points in the figure) to perform the linear fits to the [Fe/H] abundances in the same way. We obtain a slope of $-0.052\pm0.003\,\mathrm{dex/kpc}$. Previous results in the literature analyzing chemical gradients with OCs with different age and $\rgc$ ranges, show also negative trends. We include a non-exhaustive compilation of the literature results in Table~\ref{tab:lit_gradients}. 
We see no indication of flattening of the gradient at $\rgc = 10$ kpc, as suggested by several authors \citep[e.g. ][]{Twarog+1997}.

\begin{table}
\caption{Literature results of Galactocentric trends obtained with OCs.}\label{tab:lit_gradients}
\centering
\setlength\tabcolsep{2.1pt}
\begin{tabular}{|lccccccccccccc|}
 \hline
 Reference & $\mathrm{d\,[Fe/H]}/\mathrm{d\,}\rgc$ & Comment \\
 \hline
 \citet{Friel+2002}    & $-0.06\pm0.01$   & 24 OCs, $\rgc$ 7-16 kpc \\
 \citet{Jacobson+2011} & $-0.085\pm0.019$ & 10 OCs, $\rgc$ 9-13 kpc \\
 \citet{Carrera+2011}  & $-0.070\pm0.010$ & 9 OCs, $\rgc$ 6-13 kpc \\
 \citet{Reddy+2016}    & $-0.052\pm0.011$ & 28 OCs, $\rgc$ 5-12 kpc \\
 \citet{Donor+2018}    & $-0.061\pm0.004$ & 19 OCs, $\rgc$ 7-12 kpc \\
 \citet{Carrera+2019}  & $-0.052\pm0.003$ & 46 OCs, $\rgc$ 6-13 kpc \\
 \hline
\end{tabular}
\end{table}

In the same way we fit the abundance ratios [X/Fe] vs $\rgc$ of Ni, Cr, V, Sc, Ti, Si, Ca and Mg. Using the joint compilation of clusters we obtain the slopes range from $-0.005$ to $0.01\,\mathrm{dex/kpc}$, with uncertainties between $0.001$ and $0.003$ dex. We find that Ni, Cr, Si and Ca are compatible with having a flat relation, whereas for Mg the positive slope is statistically significant (just over $3\sigma$) $0.01\pm0.002\,\mathrm{dex/kpc}$. \citet{Donor+2018} also studies the trends of these for several of these elements. Remarkably, they also found a similar increasing relation for Mg with $\rgc$.
In the OCCASO sample only, the fewer number of points and the smaller range in $\rgc$ make the fits more uncertain. Comparing with the fits of the joint compilation we obtain similar results for the trends in Fe, Ca, Mg, and significantly different slopes in the cases of Ni, Cr and Si, the OCCASO ones being steeper. The slope found for Mg ($0.01\pm0.01\,\mathrm{dex/kpc}$) is very similar to the one found using the compilation, but less significant.

\subsubsection{Comparison with models}
We have compared the results of the OCCASO clusters with the pure chemical evolution model for the thin disc by \citet{Chiappini2009}, and with the chemo-dynamical simulation of the thin disc by \citet[MCM]{Minchev+2013,Minchev+2014}. The chemical evolution model provides indicative curves with the predicted chemical abundances at each $\rgc$ and age, given some assumed stellar yields and constrained mainly by solar vicinity data and HII regions gradient. It is beyond the scope of the paper to discuss in detail the nucleosynthesis of each element. The MCM simulation is a high-resolution dynamical simulation combined with the chemical evolution model of \citet{Chiappini2009}. In all plots, the abundances of both (models and simulation) are scaled such that the Solar abundance matches the most probable birth position of the Sun ($R_{\mathrm{GC}}=6$ kpc, 4.5 Gyr ago) \citep[see ][]{Minchev+2013}.

\begin{figure*}
\centering
\includegraphics[width=\textwidth]{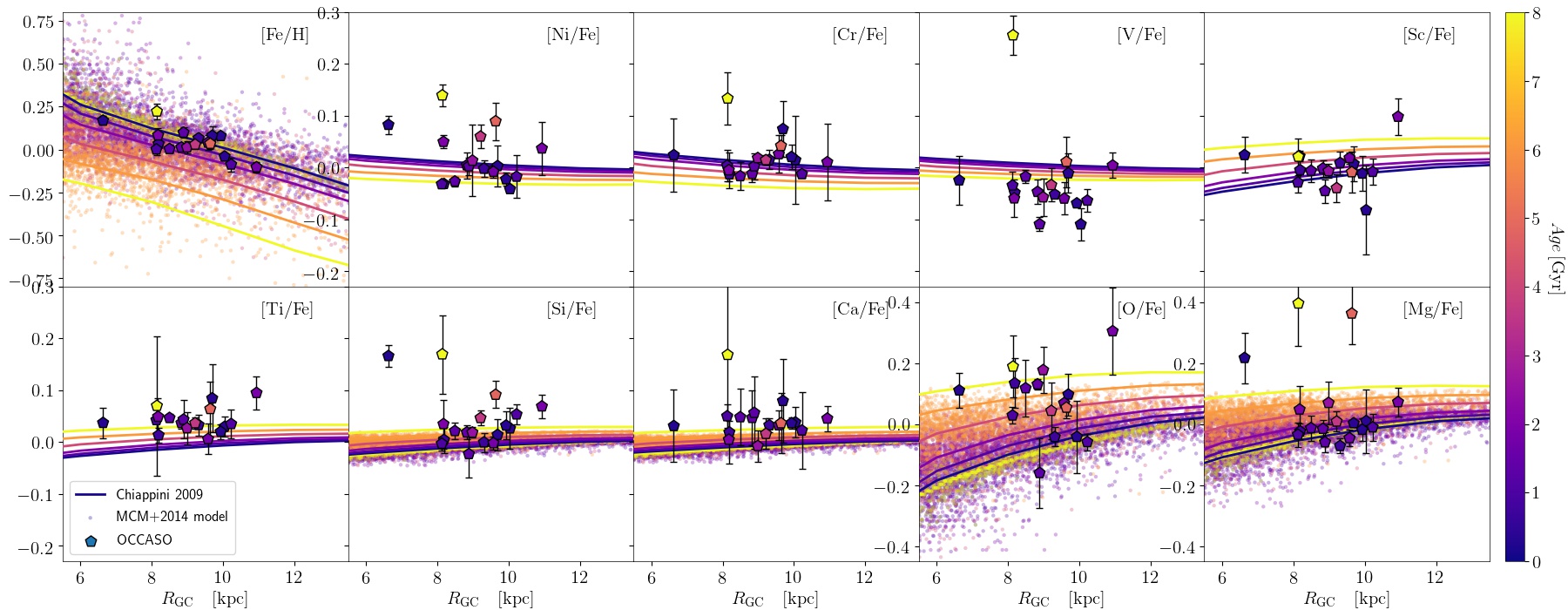}
\caption{OCCASO abundances [Fe/H] and [X/Fe] as a function of Galactocentric radius are represented as pentagons colored by age. 
Pure chemical evolution models for the thin disk by \citet{Chiappini2009} are overplotted, as well as the $N$-body chemodynamical model MCM (small dots), with the same color codes. Note the change of scale in [Fe/H], [Mg/Fe] and [O/Fe].}\label{fig:abund_RGC_mod}
\end{figure*}

The OCCASO abundances [Fe/H] and [X/Fe] vs Galactocentric radius are plotted together with the pure chemical evolution model and the MCM chemo-dynamical simulation in Figure~\ref{fig:abund_RGC_mod}.
For most of the elements the chemical evolution model predicts flat patterns and with very little changes of the gradient at different ages. Only Sc, Mg and O are sensitive to age (noticeable in the plot by the separation of the age curves), since they are mainly produced in SN type II. Therefore, when radial migration is introduced \citep{Sellwood+2002}, represented by the MCM results, it only has a consequence on the observed gradients for the chemical species sensitive to age.

In general, the curves go through the region where the clusters are. For the oldest clusters ($\gtrsim 3$ Gyr) their Galactocentric radius is probably not representative anymore of their birth position and we do not expect them to follow the predicted curves. In some cases (V, Ti and probably Ca) there are offsets between the clusters and models, even for the young clusters, this could be a matter of solar zeropoint or uncertain stellar yields. We remark the cases of Sc and Mg, where a very good agreement is seen between young OCs and models (blue points and line), but the oldest clusters do not follow anymore the curves (magenta and orange). For O the plot is more disperse because of observational errors, as already mentioned in previous sections.

\subsection{Abundances with $z$}
The existence of a vertical trend of abundance with distance to the Galactic plane has been studied in few works, e.g. in \citet[][]{Boeche+2013} using RAVE field stars. They find that abundances as a function of $z$ are mainly flat up to $\sim1.5$ kpc, where they abruptly drop for Fe, and also but more smoothly for Mg and Si. The changes at larger heights are identified with the thick disk population. To the best of our knowledge it has not been seen any significant gradient from OCs, probably because they are mainly confined in the thin disk.

\begin{figure*}
\centering
\includegraphics[width=\textwidth]{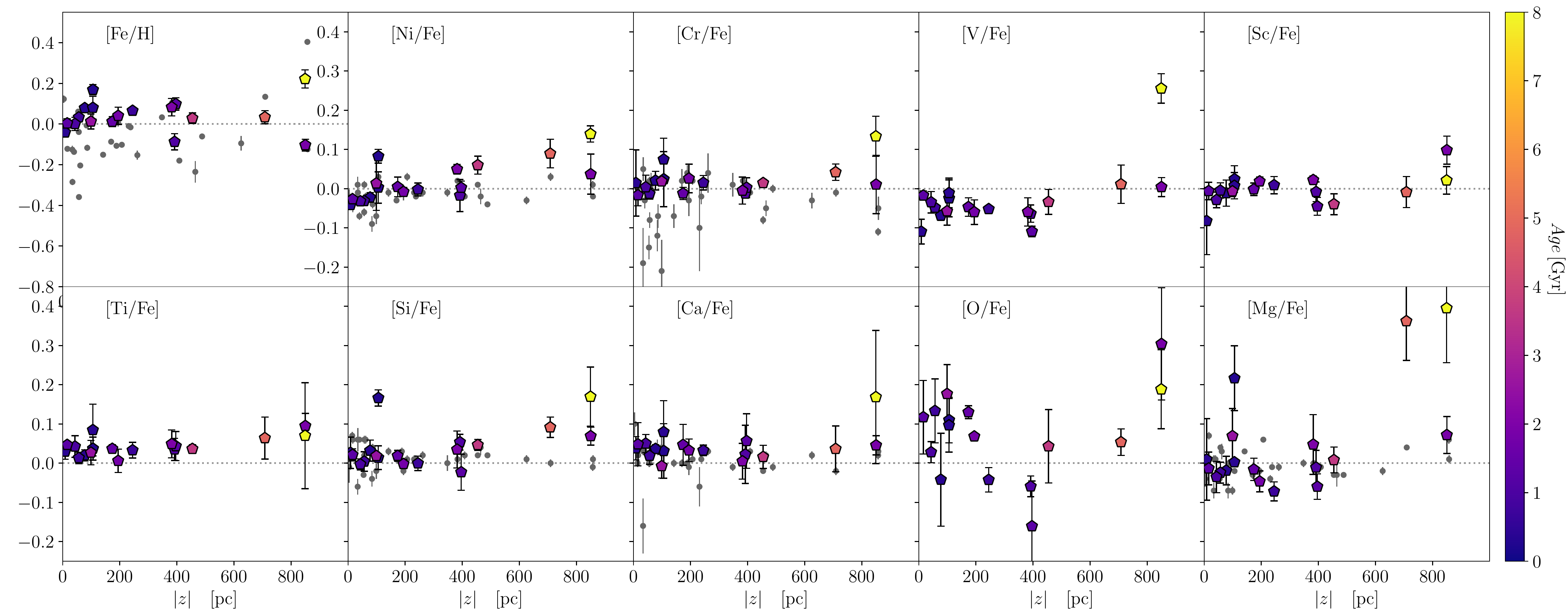}
\caption{OC abundances ratios as a function of the distance to the Galactic plane. Symbols are as in Fig.~\ref{fig:abund_RGC}. Notice the different y scale in [Fe/H].}\label{fig:abund_z}
\end{figure*}

We show the OC abundance ratios from OCCASO and from the compilation \citet{Carrera+2019} as a function of the distance to the Galactic plane in Fig.~\ref{fig:abund_z}. We do not identify any particular trend aside of the relation between $z$ and age. This is a known effect caused by the kinematics of the disc, and analyzed in detail by \citet{Soubiran+2018}. Young OCs are usually confined in the plane with low vertical velocity dispersions, while old OCs exhibit a much wider range of vertical position, and their total velocity is higher as well.

\subsection{Comparison of clusters with field stars from APOGEE}
We have used the latest APOGEE data release \citep[DR14, ][]{Abolfathi+2018} to compare the abundance trends in $\rgc$ and $z$, of OCCASO clusters with field stars. We use the estimations of distances computed from APOGEE+\emph{Gaia} DR2 data with the StarHorse code \citep{Queiroz+2018,Anders+2019}. We select from the full data set of more than 277,000 stars, thin disc stars by limiting the height above the plane $|z|$ up to 1 kpc, and mainly red giants with $1.5<\log g<3.5$. We end up with a sample of $\sim99,000$ stars. We investigate only Fe, Ni, Cr, Si, Ca and Mg the same as for the sample of \citet{Carrera+2019}. We do not attempt to perform any fit since field stars have less reliable ages and suffer migration as well.

\begin{figure}
\centering
\includegraphics[width=0.5\textwidth]{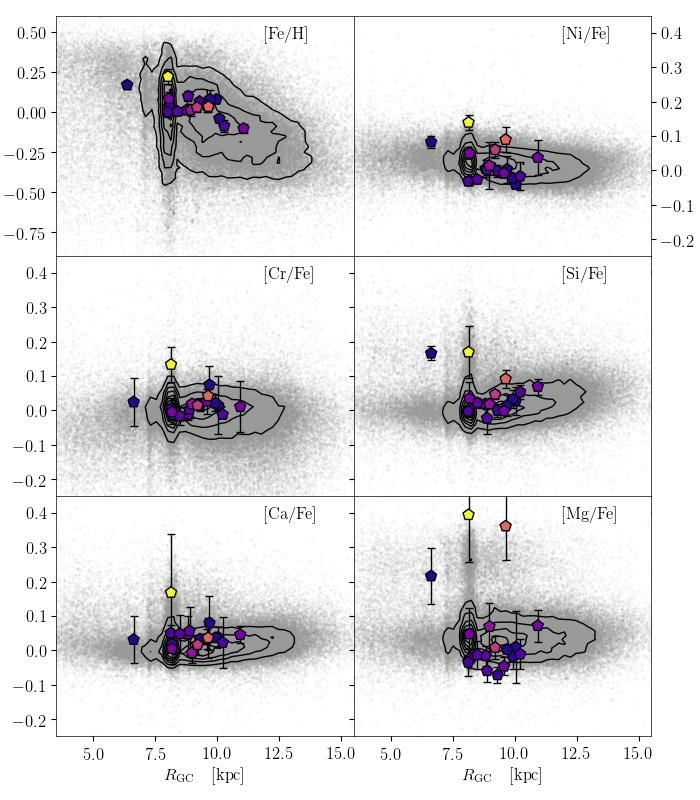}
\caption{[Fe/H] and abundance ratios [X/Fe] as a function of $\rgc$ of OCCASO clusters (pentagons color coded by age, as in Fig.~\ref{fig:abund_RGC_mod}, and field stars from APOGEE (see text). Solid lines represent the density contours of the APOGEE field stars. Notice the different y scale for [Fe/H].}\label{fig:field_OCs_rgc}
\end{figure}

We plot abundances [Fe/H] and [X/Fe] as a function of $\rgc$ in Fig.~\ref{fig:field_OCs_rgc}. The selection function of APOGEE can be clearly seen in this figure, as vertical stripes and an overdensity in the solar vicinity. The clusters follow the general distribution of the field stars in all elements, except for Mg, where a slight offset can be seen between the peak of the field distribution and most of the clusters.
We can see that there is significant field population in the position of all clusters, even for those clusters which appear as outliers in several elements (Ni, Si, Mg). Especially, for Mg vs $\rgc$ an overdensity is seen for the field stars at Mg$>0.2$, probably thick disc contribution, in the same region where the three outlier cluster are. The significant increasing trend seen in [Mg/Fe] in Fig.~\ref{fig:abund_RGC} does not seem evident from the field stars.

\begin{figure}
\centering
\includegraphics[width=0.5\textwidth]{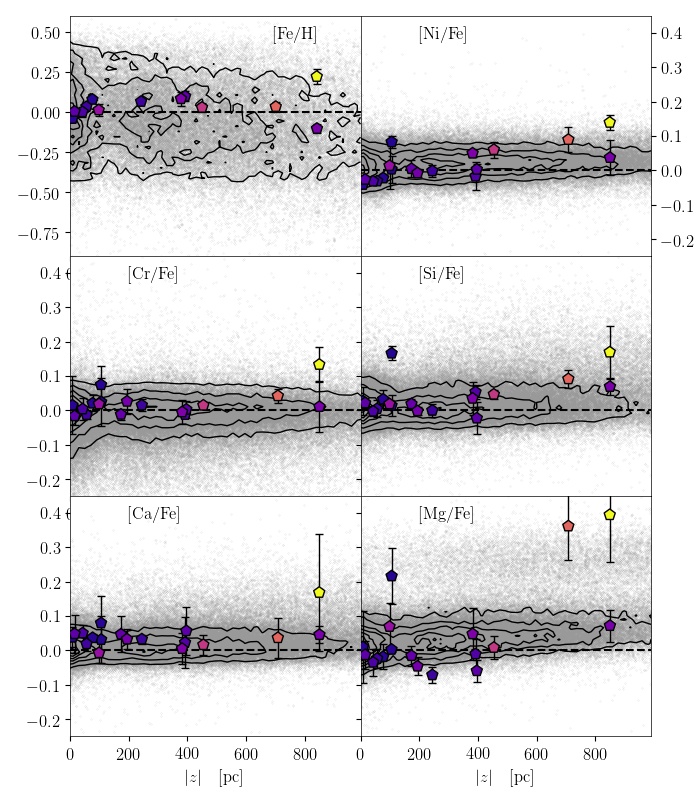}
\caption{[Fe/H] and abundance ratios [X/Fe] as a function of $|z|$ of OCCASO clusters (pentagons color coded by age, as in Fig.~\ref{fig:abund_RGC_mod}, and field stars from APOGEE (see text). Solid lines represent the density contours of the APOGEE field stars. Notice the different y scale for [Fe/H].}\label{fig:field_OCs_z}
\end{figure}

We plot the same abundances as a function of $|z|$ in Fig.~\ref{fig:field_OCs_z}. Again, the general trend from field stars is followed by clusters. In the case of Mg we observe an increasing trend seen towards high $z$ for the field stars, which is not so clear for the clusters.

\subsection{Abundances with $Age$}
One of the most clear advantages of using clusters to study the chemical evolution of the Galaxy is that ages are much better determined compared to field stars. However, they provide a biased age distribution towards younger clusters. In this section we use the ages from Table~\ref{tab:clusters}.

\begin{figure*}
\centering
\includegraphics[width=\textwidth]{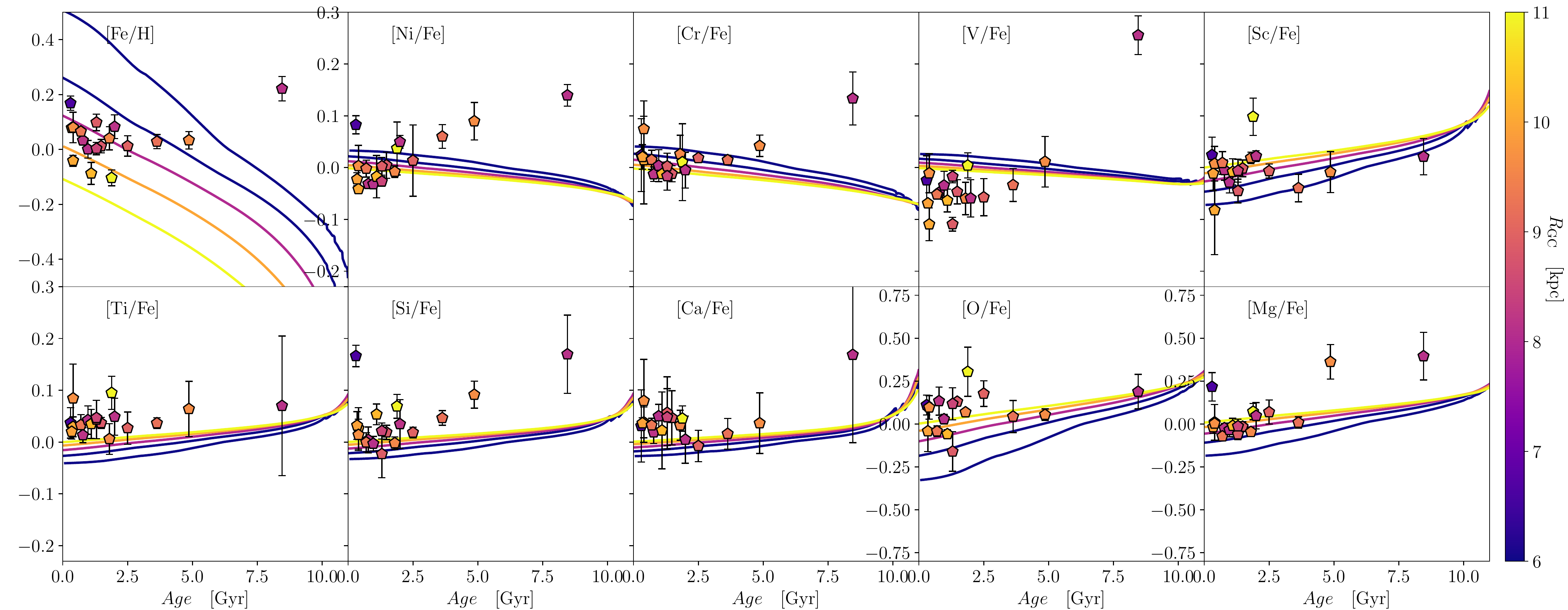}
\caption{OC abundance ratios and pure chemical evolution model predictions as a function of age. The colors correspond to the Galactocentric radii.}\label{fig:abund_age}
\end{figure*}

In Fig.~\ref{fig:abund_age} we plot [Fe/H] and the abundance ratios [X/Fe] of the different elements as a function of the cluster age, and colored by Galactocentric radius. We also include the chemical evolution model prediction \citep{Chiappini2009} for five different Galactocentric radii.
The age distribution of our cluster sample is more concentrated at ages younger than 3 Gyr, because there are very few old clusters in our Galaxy. Also, one of the oldest clusters is NGC~6791, a known chemically peculiar OC already mentioned in Sec. 4.

We expect to see scatter in the age-[Fe/H] relation in the particular age range we have most of the clusters, because of radial mixing \citep[see discussion in ][]{Minchev+2013}.

\begin{itemize}
	\item For the Fe-peak elements Ni, Cr and V, a flat relation is expected from the models. We obtain however large slopes for Ni and V. The increasing trends for these two elements seem to be also followed by our two oldest clusters up to 8 Gyr. However, as already mentioned we have poor statistics at ages older than 3 Gyr.
	\item Ti and Ca seem to follow better the model trends. For Si we see a increasing trend (mainly driven by the oldest clusters) with small errors in our abundances, while the models predict a flatter trend.
	\item For Mg, O and Sc the chemical models predict slopes slightly steeper than for the other elements. The observational picture seems to be compatible with the predictions. These elements are the most sensitive to age and radial migration, seen by the separation between the age curves in Fig.~\ref{fig:abund_age}.
\end{itemize}

To understand the differences there must be kept in mind that there is room for improvement of the chemical evolution models, especially after Gaia DR2, distances and ages are orders of magnitude more precise than those available before.

Our age range and sample size is smaller than other samples that have been used to analyze age trends, usually done with field stars in the solar neighbourhood, e.g. recently by \citet[][]{DelgadoMena+2019,Bedell+2018}. For Fe peak elements, \citet{Bedell+2018} obtain similar flat and increasing relations, respectively for Cr and Sc. For V and Ni our trends seem much steeper than those of \citet{Bedell+2018}. Regarding the $\alpha$ elements, our results fit in the general picture seen by both studies with slightly steeper slopes for Mg and O, and less steep relations for Si, Ca and Ti.

\section{Conclusions}
In this work we have derived abundances of 10 chemical species including Fe-peak and $\alpha$ elements (Fe, Ni, Cr, V, Sc, Ti, Si, Ca, O, Mg) for a sample of 115 stars in 18 OCs as part of the OCCASO survey. With this data we have investigated the Galactic trends with $\rgc$, $z$ and age, and compared them with the predictions of a pure chemical model and a chemo-dynamical simulation.

We extensively compare our results of abundances with the literature finding no global systematics of our measurements. We compute cluster abundances averaging all bona fide member stars, constrained with the help of \emph{Gaia} DR2 memberships. Our cluster dispersions in [X/Fe] are in general lower than 0.05 dex, with exception of some clusters (e.g. NGC~6791) in some elements (e.g. Sc, O),due to the higher uncertainties in measuring their abundances.

We investigate the Galactic trend of abundances as a function of Galactocentric radius. We use OCCASO clusters together with the compilation by \citet{Carrera+2019} of APOGEE and GALAH OCs with more than 4 observed stars. The compatibility of the two samples, and the large covered $R_{\mathrm{GC}}$, allows us to investigate the trends in the different elements. We first have computed the present day metallicity gradient using only the youngest clusters obtaining $-0.05\pm0.01\,\mathrm{dex/kpc}$ and $-0.06\pm0.01\,\mathrm{dex/kpc}$, for age$<2$ Gyr (14 OCs) and age$<1.5$ Gyr (11 OCs), respectively. These are compatible values as those computed by \citet{Genovali+2014} using Cepheids. We use the full compilation including clusters of all ages to investigate trends in the different elements. We find the slopes: $-0.051\pm0.003$ [Fe/H], $0.002\pm0.001$ [Ni/Fe], $-0.005\pm0.003$ [Cr/Fe], $0.005\pm0.002$ [Si/Fe], $0.002\pm0.001$ [Ca/Fe], $0.010\pm0.002$ dex kpc$^{-1}$ [Mg/Fe]. We find a significant positive slope only for [Mg/Fe], as the chemical models predict. A very similar slope but with larger uncertainties is found when we perform a fit to the OCCASO sample only.

We compare our results to the predictions of the pure chemical evolution model by \citet{Chiappini2009} for the different elements as a function of $\rgc$. A good compatibility is seen for the youngest clusters in most elements (except an offset observed in [V/Fe] and [Ca/Fe]). The older ones ($\gtrsim3\,\mathrm{Gyr}$) deviate from the predictions of the model for all elements, probably due to the effect of radial migration: current Galactocentric radii for old clusters are not representative of their birth positions and we do not expect them to follow the models. This effect is shown by the MCM chemo-dynamical simulation \citep{Minchev+2013,Minchev+2014}.

We investigate the OCs abundances as a function of the distance to the Galactic plane. We do not find any clear trend.

We use the sample of APOGEE DR14 stars, with distances were computed by \citet{Queiroz+2018,Anders+2019} using Gaia DR2, to compare the general trends in $\rgc$ and $|z|$ of  clusters and field stars. We have seen that the general trends of the clusters are followed by the field distribution. There is significant field population in the position of all clusters, even for those clusters which appear as outliers in several elements (Ni, Si, Mg). In the case of Mg as a function of $|z|$, we find an increasing trend for the field stars of APOGEE, not clearly shown by the clusters.

We also investigate the relation of abundance ratios with age drawn by OCCASO clusters, and compared with the chemical evolution model. The clusters are very concentrated towards ages younger than 3 Gyr. We see a significant scatter in the age-[Fe/H] relation in this particular age range, generally understood as the effect of radial mixing. A good agreement is seen for Sc, Mg and O, elements more sensitive to age and radial migration. We find that in some elements there are large differences between the model predictions and the obtained trends (Ni, Cr, V and Si), showing that there is still room for improvement in the chemical evolution models.

\section*{Acknowledgements}
We thank the referee for her/his comments. We greatfully thank Ulrike Heiter and Javier Olivares for their useful discussions and help.

This work is based on observations made with the Nordic Optical Telescope, operated by the Nordic Optical Telescope Scientific Association, and the Mercator Telescope, operated on the island of La Palma by the Flemish Community, both at the Observatorio del Roque de los Muchachos, La Palma, Spain, of the Instituto de Astrof\'isica de Canarias. This work is also based on observations collected at the Centro Astron\'omico Hispano en Andaluc\'ia (CAHA) at Calar Alto, operated by the Instituto de Astrof\'isica de Andaluc\'ia (CSIC).

This research made extensive use of the SIMBAD database, and the VizieR catalogue access tool, operated at the CDS, Strasbourg, France, and of NASA Astrophysics Data System Bibliographic Services.
This work has made use of results from the European Space Agency (ESA) space mission Gaia, the data from which were processed by the Gaia Data Processing and Analysis Consortium (DPAC). Funding for the DPAC has been provided by national institutions, in particular the institutions participating in the Gaia Multilateral Agreement.

L.C. acknowledge support from the "programme national cosmologie et galaxies" (PNCG) of CNRS/INSU.
CJ, LBN and JCH acknowledge support by the MINECO (Spanish Ministry of Economy) through grant ESP2016-80079-C2-1-R (MINECO/FEDER, UE) and MDM-2014-0369 of ICCUB (Unidad de Excelencia "Mar\'ia de Maeztu").
CG acknowledges financial support through the grant (AEI/FEDER, UE) AYA2017-89076-P, as well as by the Ministerio de Ciencia, Innovaci\'on y Universidades (MCIU), through the State Budget and by the Consejer\'\i a de Econom\'\i a, Industria, Comercio y Conocimiento of the Canary Islands Autonomous Community, through the Regional Budget.

\bibliographystyle{mnras}
\bibliography{biblio_v4,biblio_linelist}

\newpage
\appendix

\section{Detailed cluster abundances plots}\label{sec:abplots}

Star abundances [X/Fe] as a function of [Fe/H] for each cluster and element analyzed are shown.

\begin{figure*}
\centering
\includegraphics[width=0.8\textwidth]{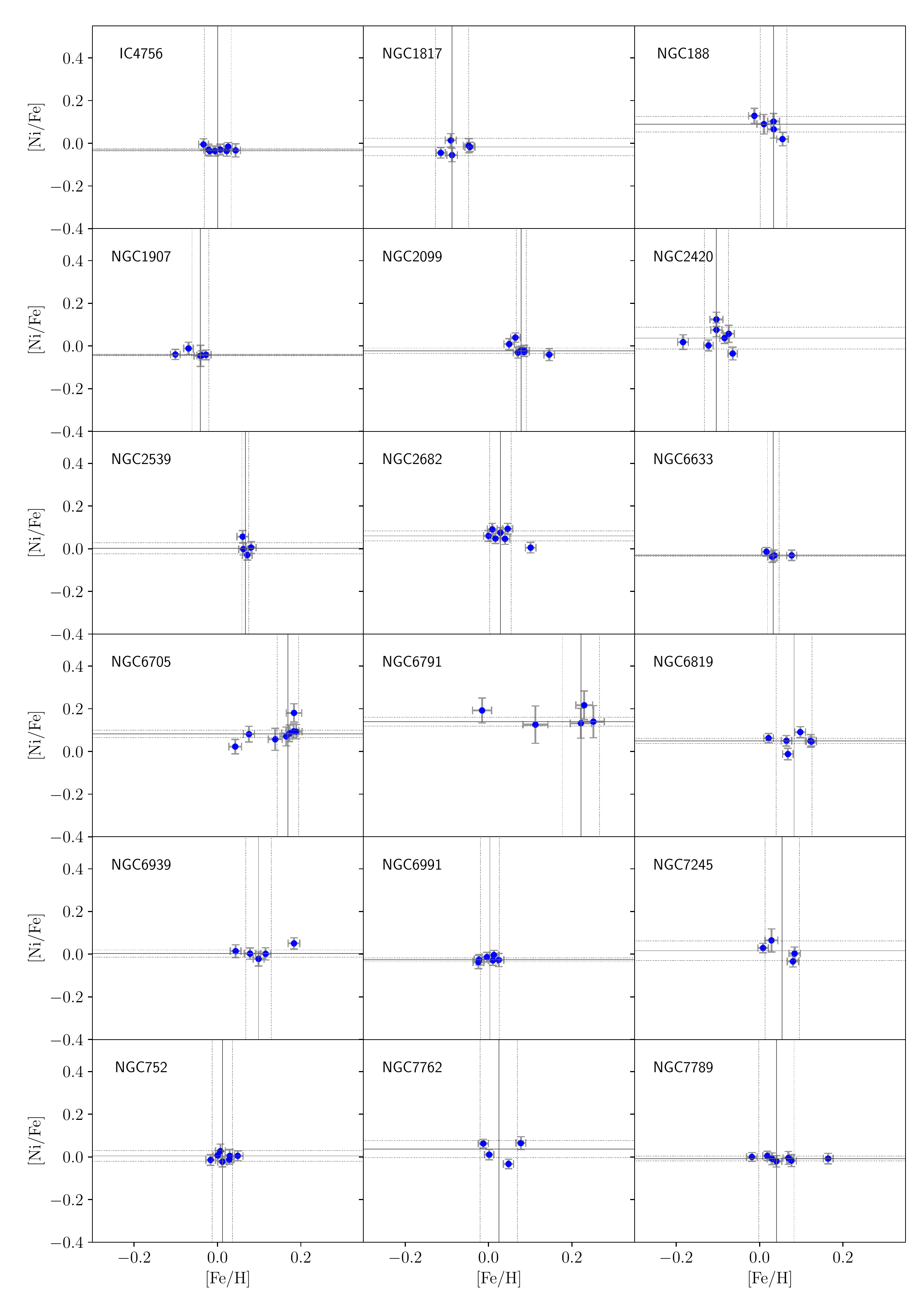}
\caption{[Ni/Fe] abundance for the $18$ studied OCs as a function of Fe abundance. The black solid and dashed lines indicate the median and $1MAD$ level. Stars detected as non members or spectroscopic binaries in Paper I and II, are not plotted here and have not been used for the final OC abundance.}\label{fig:clusters_NiFe}
\end{figure*}

\begin{figure*}
\centering
\includegraphics[width=0.8\textwidth]{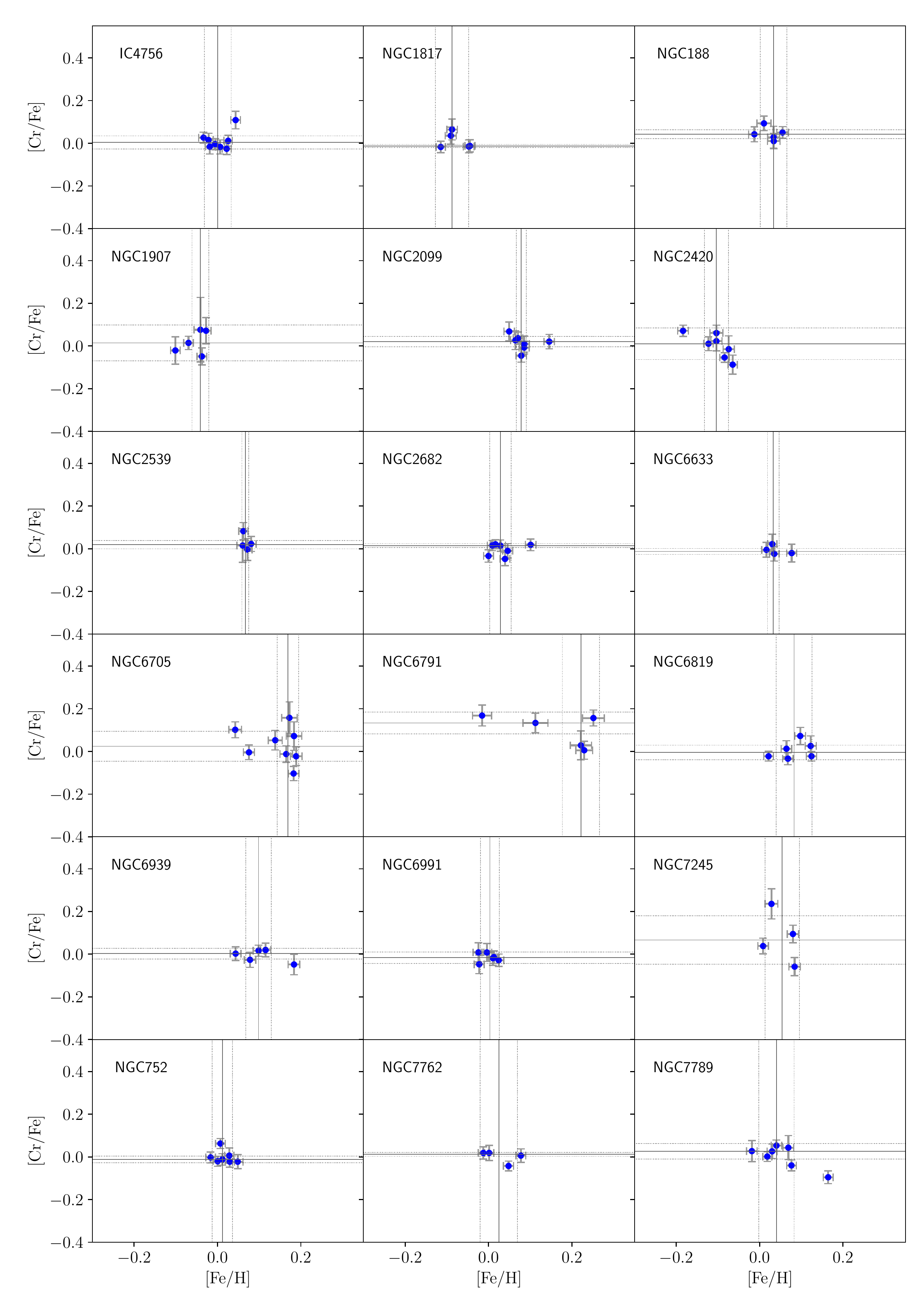}
\caption{As in \ref{fig:clusters_NiFe}, for Cr.}\label{fig:clusters_CrFe}
\end{figure*}

\begin{figure*}
\centering
\includegraphics[width=0.8\textwidth]{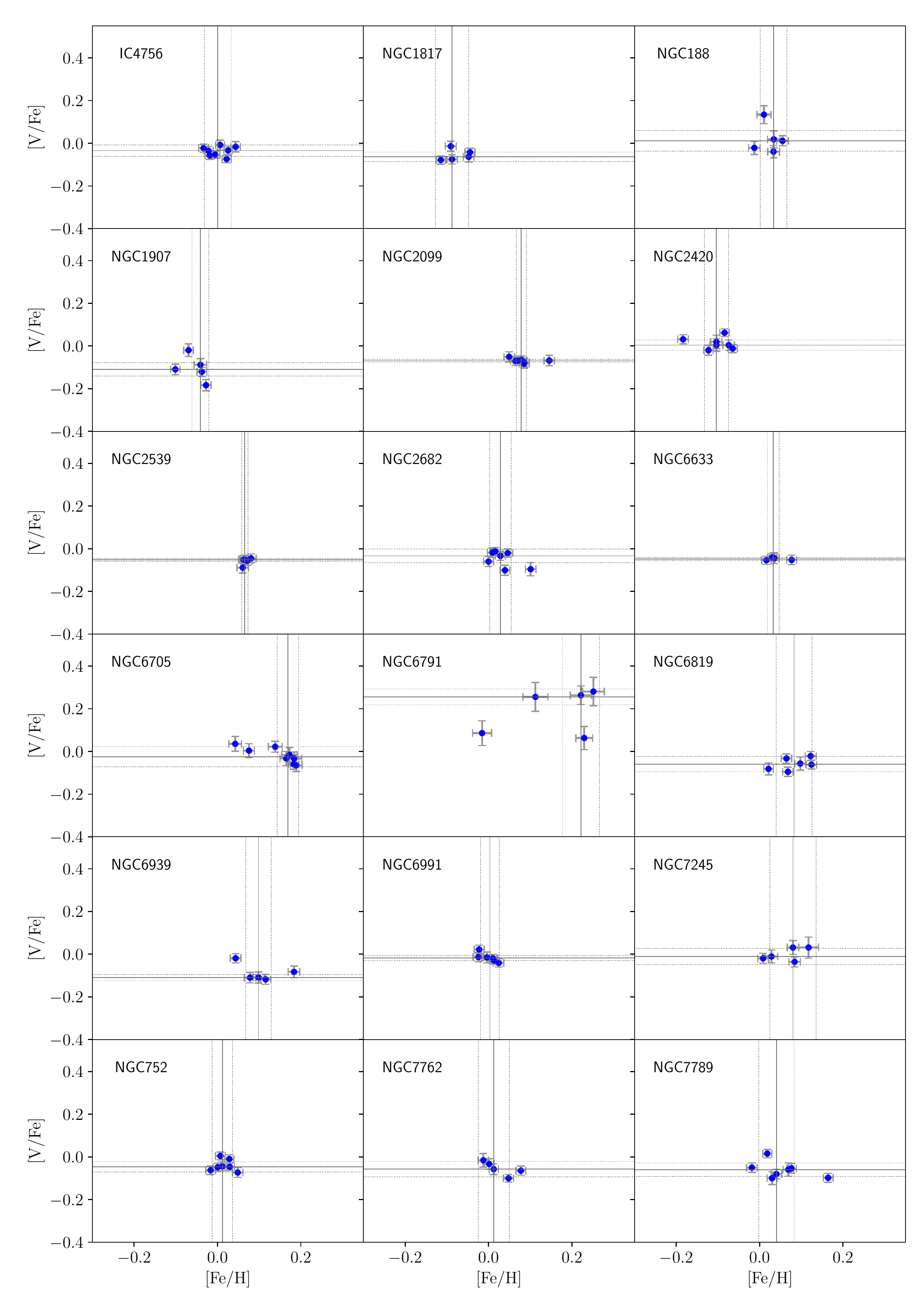}
\caption{As in \ref{fig:clusters_NiFe}, for V.}\label{fig:clusters_VFe}
\end{figure*}

\begin{figure*}
\centering
\includegraphics[width=0.8\textwidth]{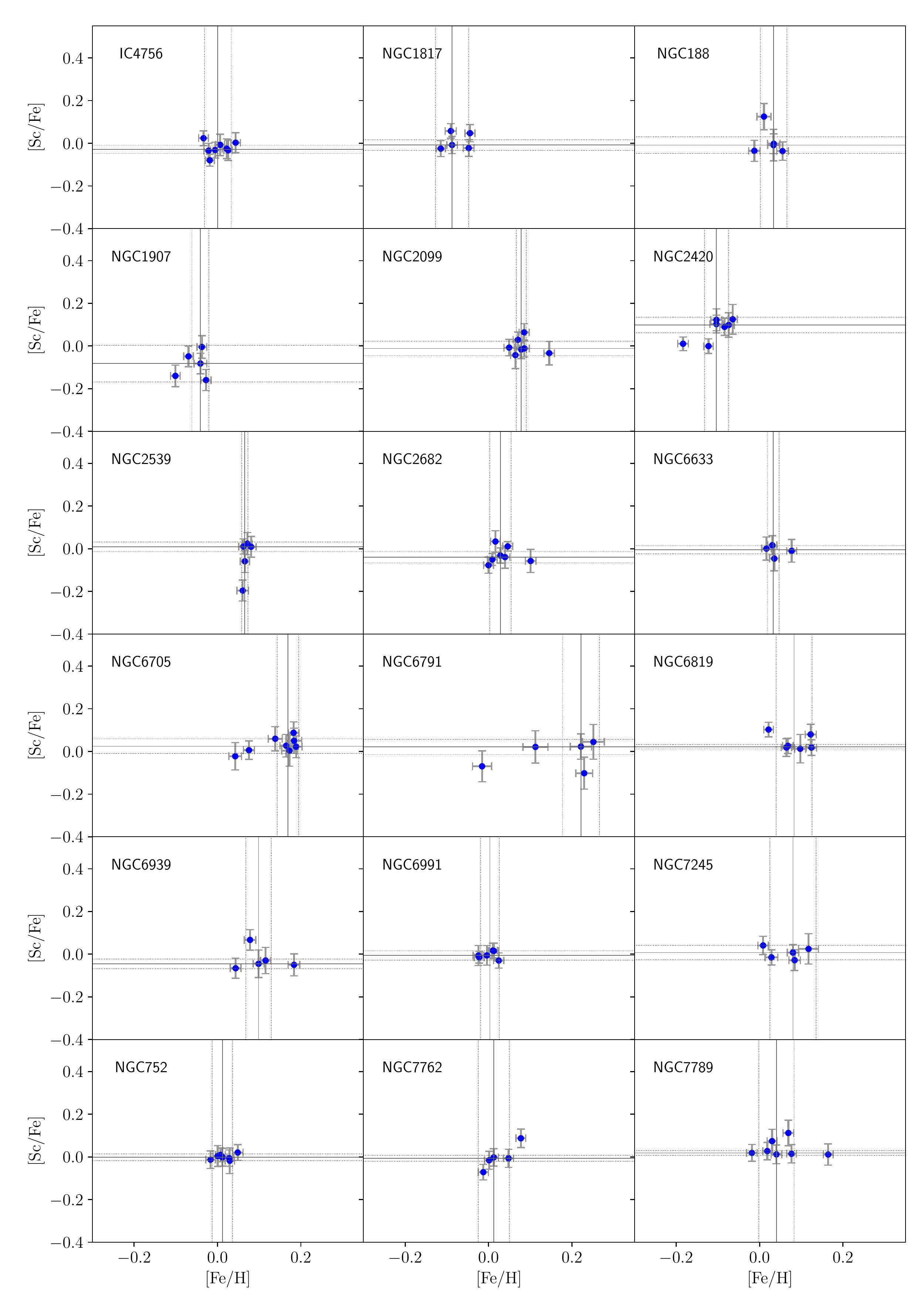}
\caption{As in \ref{fig:clusters_NiFe}, for Sc.}\label{fig:clusters_ScFe}
\end{figure*}

\begin{figure*}
\centering
\includegraphics[width=0.8\textwidth]{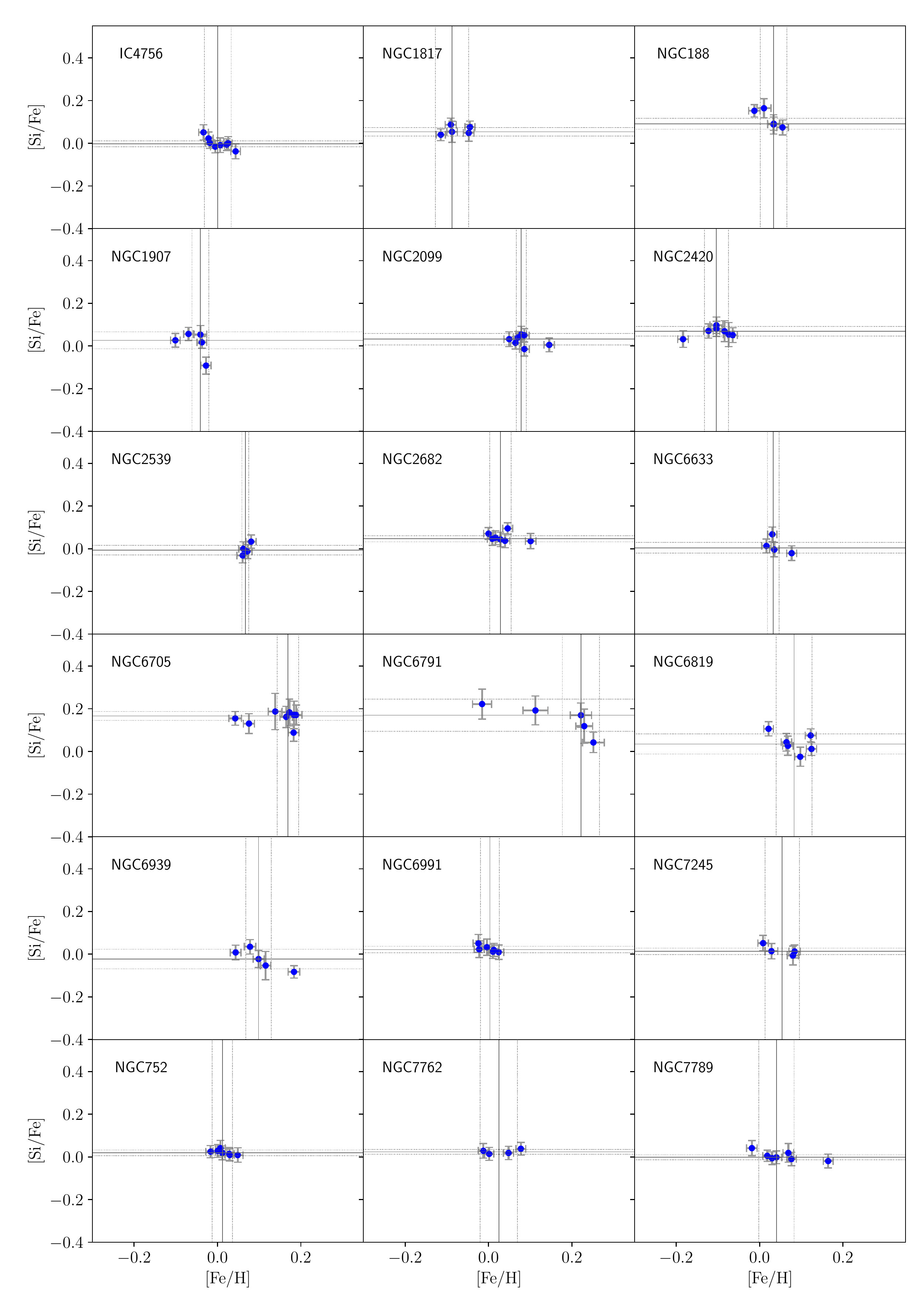}
\caption{As in \ref{fig:clusters_NiFe}, for Si.}\label{fig:clusters_SiFe}
\end{figure*}

\begin{figure*}
\centering
\includegraphics[width=0.8\textwidth]{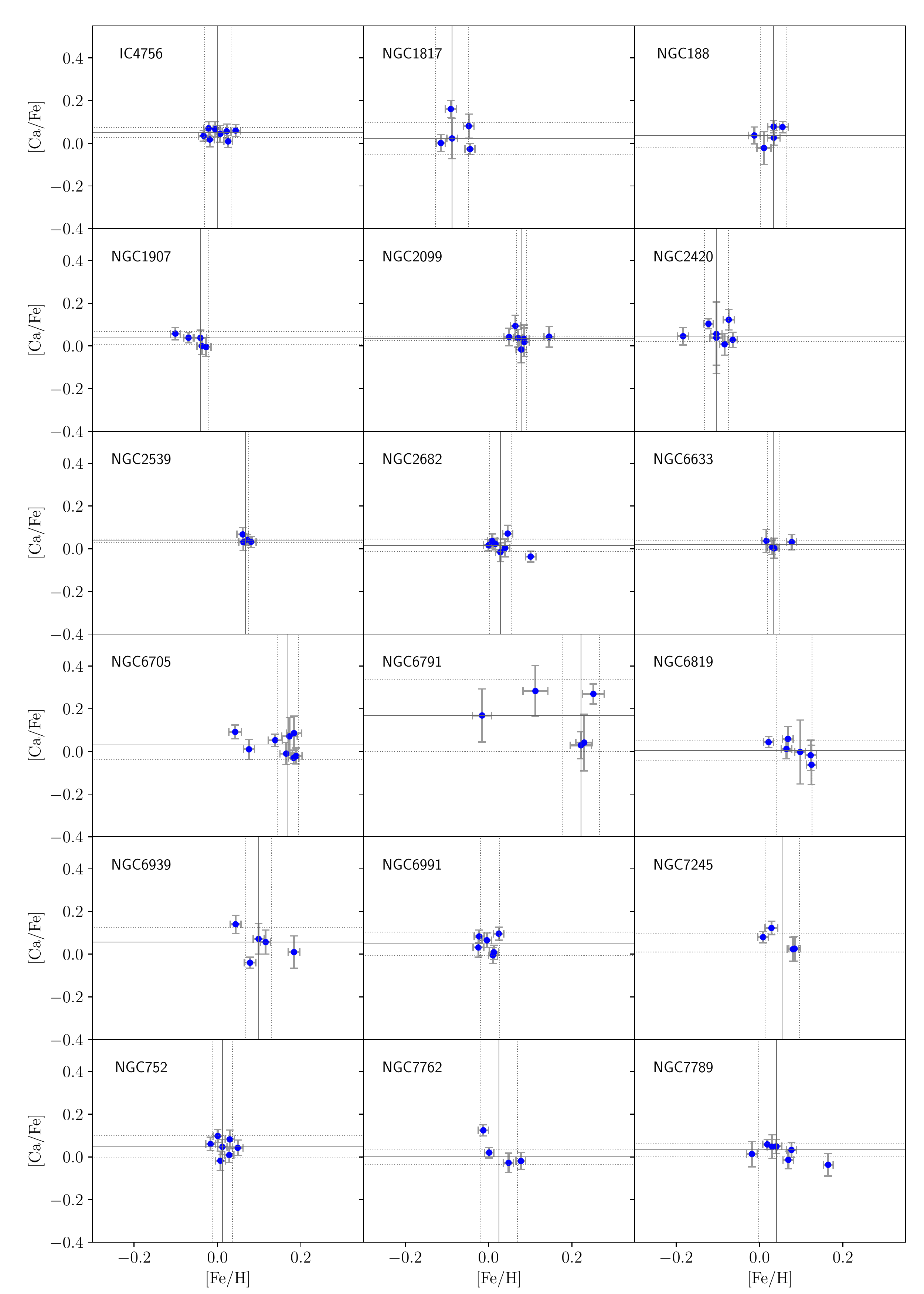}
\caption{As in \ref{fig:clusters_NiFe}, for Ca.}\label{fig:clusters_CaFe}
\end{figure*}

\begin{figure*}
\centering
\includegraphics[width=0.8\textwidth]{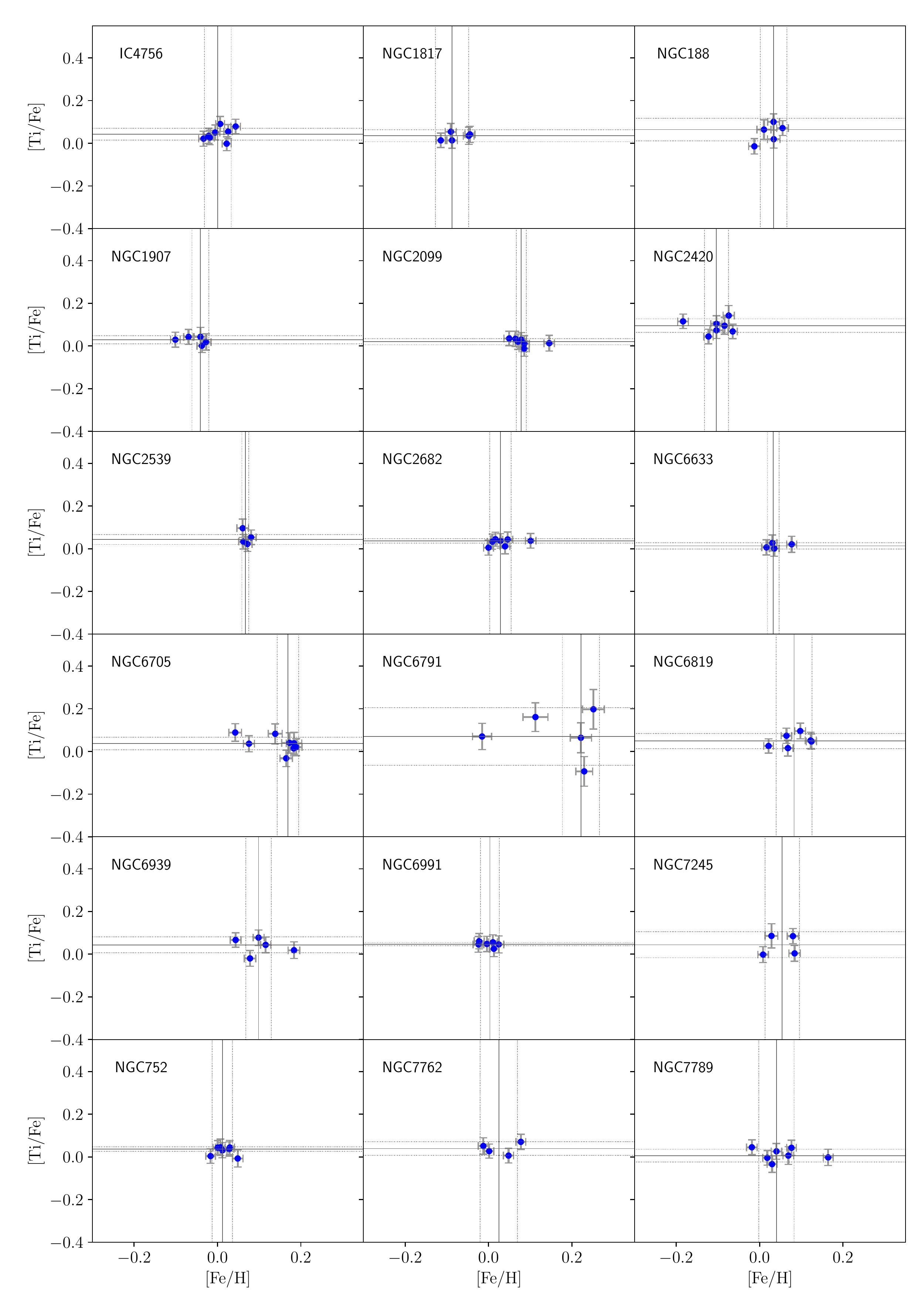}
\caption{As in \ref{fig:clusters_NiFe}, for Ti.}\label{fig:clusters_TiFe}
\end{figure*}

\begin{figure*}
\centering
\includegraphics[width=0.8\textwidth]{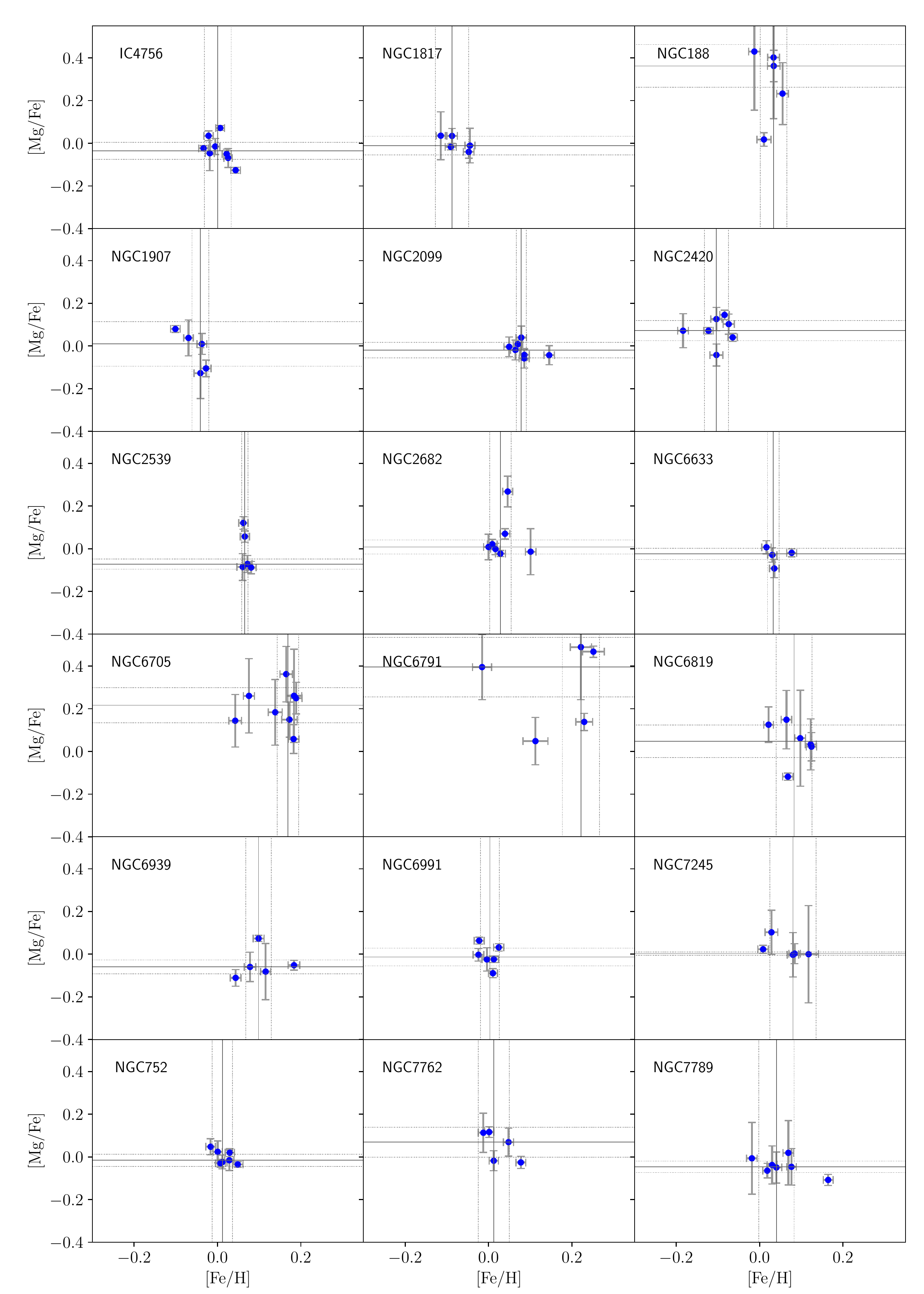}
\caption{As in \ref{fig:clusters_NiFe}, for Mg.}\label{fig:clusters_MgFe}
\end{figure*}

\begin{figure*}
\centering
\includegraphics[width=0.8\textwidth]{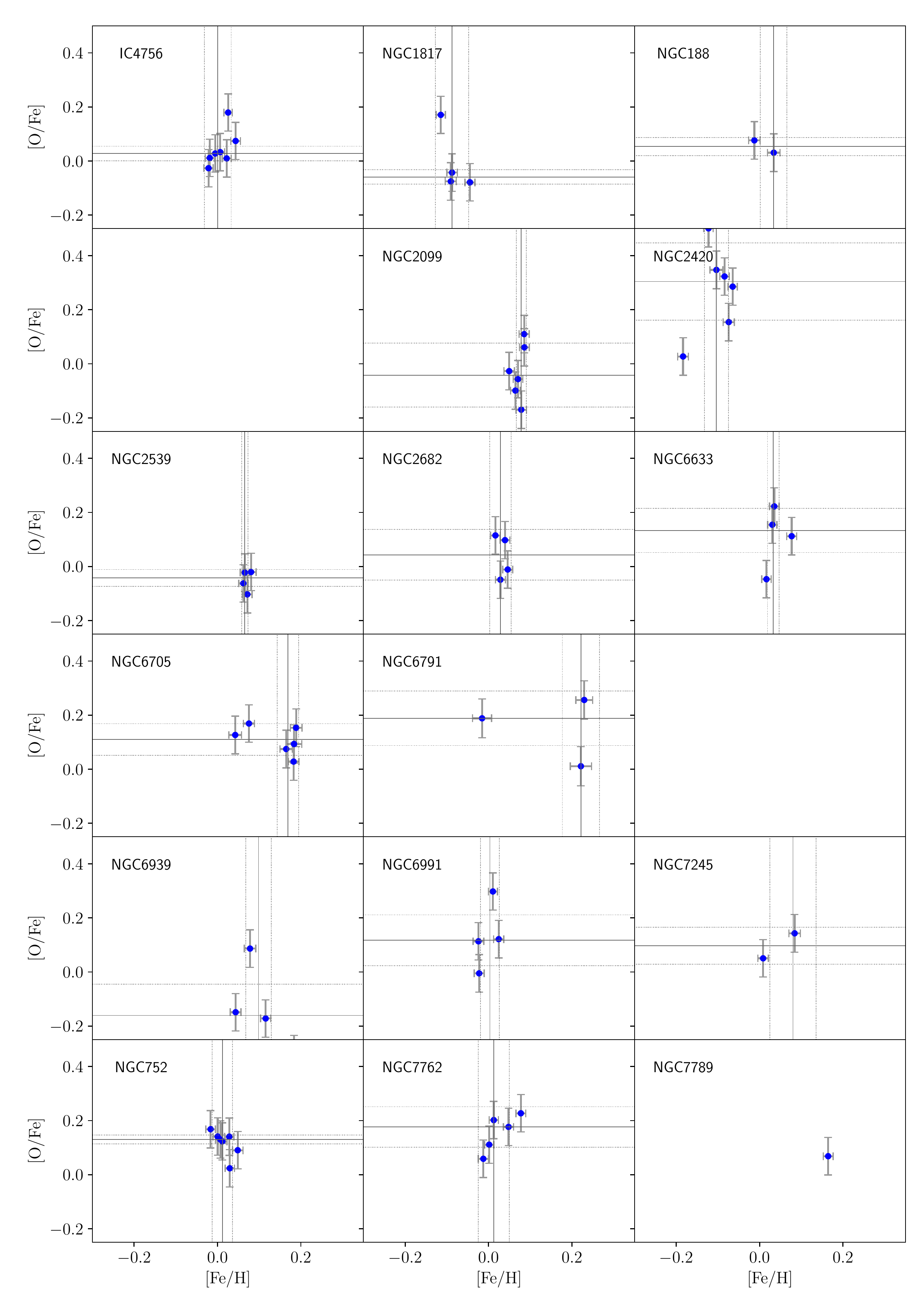}
\caption{As in \ref{fig:clusters_NiFe}, for O. Oxygen abundance could not be measured in any star of the clusters NGC~1907 and NGC~6819. For NGC~7789 only one star has a value of oxygen.}\label{fig:clusters_OFe}
\end{figure*}

\bsp	

\label{lastpage}
\end{document}